\documentstyle[preprint,aps]{revtex}
\begin{document}
\title{Classical and Quantum Chaos in a quantum dot\\ in time-periodic
magnetic fields}
\author{R. Badrinarayanan and Jorge V. Jos\'e}
\address{{\it Department of Physics and Center for Interdisciplinary
Research on Complex Systems,\\ Northeastern University, Boston
Massachusetts 02115, USA}} 
\maketitle
\begin{abstract}
We investigate the classical and quantum dynamics of an electron 
confined to a circular quantum dot in the presence of 
homogeneous  $B_{dc}+B_{ac}$ magnetic fields. The classical motion 
shows a transition to chaotic behavior depending on the ratio 
$\epsilon=B_{ac}/B_{dc}$ of field magnitudes and the cyclotron 
frequency ${\tilde\omega_c}$ in units of the drive frequency. 
We determine a phase boundary between regular and chaotic 
classical behavior in the $\epsilon$ vs ${\tilde\omega_c}$ plane. 
In the quantum regime we evaluate the quasi-energy spectrum of the 
time-evolution operator. We show that the nearest neighbor 
quasi-energy eigenvalues show a transition from level 
clustering to level repulsion as one moves from the regular to 
chaotic regime in the $(\epsilon,{\tilde\omega_c})$ plane. 
The $\Delta_3$ statistic confirms this transition. In the chaotic 
regime, the eigenfunction statistics coincides with the 
Porter-Thomas prediction. Finally, we explicitly establish the 
phase space correspondence between the classical and quantum 
solutions via the Husimi phase space distributions of 
the model. Possible experimentally feasible conditions to see 
these effects are discussed.

Pacs: 05.45.+b
\end{abstract}
\maketitle
\section{Introduction}
\label{sec:intro}
In this paper, we present results of a 
study of the behavior of an electron confined to a disk of finite 
radius, subjected to spatially uniform, constant ($B_{dc}$) plus 
time-varying ($B_{ac}$) perpendicular magnetic fields. This allows 
us to analyze an old problem which exhibits some very novel behavior 
because of the time-dependent field.  Without this time 
varying component of the field, the electronic states form the  
oscillator-like Landau levels\cite{fock}. With the 
addition of confinement, this constant field problem was studied in 
great detail by Dingle\cite{dingle}. 
He obtained perturbative solutions and subsequently others 
obtained numerical and exact\cite{robnik} solutions.  The 
solutions depend on the ratio of the cyclotron radius $\rho_c$ to the 
confinement radius $R_0$. One of the most important consequences of 
confinement is the presence of  `skipping orbits', which play an 
important role, for example, in the Quantum Hall 
Effect\cite{prange}.  

This problem is of significant interest as a consequence of two 
independent  developments over the past few years. One, the 
important advances in our knowledge of classical chaos\cite{ll}, 
and to a lesser extent, it's quantum and semiclassical 
counterparts\cite{casati1}; and two, the 
spectacular advances in the fabrication of very clean mesoscopic 
quantum devices\cite{beenakker}, where a high-mobility 
two-dimensional electron gas is trapped within 
a boundary of controlled shape. We attempt  
to begin to bring the two fields together by asking how 
this model system behaves from the classical dynamical point of 
view and what it's quantum signatures are.
We predict ranges of fields and frequencies where some novel 
effects may be experimentally observable. In this paper, we consider the 
single-electron case and leave for a future publication the many electron 
problem.

This paper is organized as follows: In section II we define the model 
with its classical and quantum-mechanical properties, 
elucidate the important parameters in the 
problem and describe the general method of solution. In section 
III, we investigate the properties of the classical model. Based on a 
combination of analytic and numerical analysis, we obtain a 
`phase diagram' in the parameter space of the 
system, which separates the quasi-integrable from the chaotic 
regions. This phase diagram is shown in Fig.1.  The vertical axis is 
the ratio $\epsilon=B_{ac}/B_{dc}$ of the magnitudes of the fields, 
and the horizontal axis is the Larmor frequency normalized to the 
{\it a.c.} drive frequency, ${\tilde\omega_c}=\omega_c/\omega_0$. 
This phase diagram is of paramount importance in making the connection between 
the classical and quantum solutions. The values of the d.c. field 
$B_{dc}$ and drive frequency $\omega_0$ depend on the radius of the 
dot $R_0$ and certain other parameters. However, to give an idea of the 
magnitudes of the physical parameters involved, let us pick two 
representative points on the diagram: $({\tilde{\omega_c}},\epsilon)$ = 
(0.1, 0.1) corresponds to $\omega_0$ = 20 GHz and $B_{dc}$ = 20 gauss 
when $R_0 = 1\mu m$, while $\omega_0$ = 800 MHz and $B_{dc}$ = 0.08 gauss for 
$R_0 = 5\mu m$.  Similarly,  $({\tilde{\omega_c}},\epsilon)$ = 
(2.0, 2.0) corresponds to $\omega_0$ = 20 GHz and $B_{dc}$ = 800 gauss 
for $R_0 = 1\mu m$, while $\omega_0$ = 20 GHz and $B_{dc}$ = 32 gauss for 
$R_0 = 5\mu m$. The details of the these estimates are 
presented in Section V.

We analytically obtain conditions and 
look at various kinds of fixed points of the classical solutions. In 
section IV we study the spectral statistics of the quantum evolution 
operator, which shows clear signatures of the classical transition 
from quasi-integrabality to chaos. 
We also discuss the eigenfunctions properties  in different 
regimes using the $\chi ^2$ distribution of $\nu$ freedoms as a 
convenient parameterization of the results. 
Then, we turn to semiclassical correspondences, where we 
use a phase-space approach to the quantum eigenfunctions, and make 
direct connections with various types of classical phase space 
periodic orbits. In section V we discuss possible experimental 
scenarios where the predicted effects may be observable. 
Finally, in section VI we summarize our results 
and present our conclusions.  
\section{The Model}
\label{sec:model}
The model of a quantum dot we consider here is that of an
electron confined to a disk of radius $R_0$  subject  to 
steady ({\it d.c.}) and time-periodic ({\it a.c.}) magnetic fields. 
Choosing the cylindrical gauge, where the vector potential 
${\bf A}(\vec \rho,t) 
= {1\over 2}B(t)\, \rho\, \hat e_\phi$, $B(t)$ being the time-dependent magnetic 
field, the 
quantum mechanical single-particle 
Hamiltonian in the coordinate representation is given by 
\begin{equation}
\label{eq:a}
H = -\frac{{\hbar^2}}{2m^*}\left( \frac{d^2}{d\rho^2} + 
\frac{1}{\rho}\frac{d}{d\rho} + 
\frac{1}{\rho^2}\frac{d^2}{d\phi^2} \right) + \frac{1}{8} m^* 
\Omega^2(t) \rho^2 + \frac{1}{2} \Omega(t) L_z, \quad 0 \leq 
\rho \leq R_0,
\end{equation}
where $m^*$ is the effective mass of the electron (roughly 0.067$m_e$ in 
GaAs-AlGaAs semiconductor quantum dots) \cite{beenakker},
$L_z$ is the operator of 
the conserved angular momentum, and 
$\Omega(t) = e^{*}B(t)/m^*c$, $e^{*}$ and $c$ 
being the effective electronic charge ($e^{*}\sim 0.3e$) and the 
speed of light, respectively. 
Let the magnetic field be of the form, $B(t) = B_{dc} + 
B_{ac} f(t)$, where $f(t)=f(t+T_0)$ is some periodically time 
varying function. 
We can separate the Hamiltonian $H=H_{dc} + H_1(t)$, where
\begin{equation}
\label{eq:b}
H_{dc} = -\frac{{\hbar^2}}{2m^*}\left( \frac{d^2}{d\rho^2} 
+ \frac{1}{\rho}\frac{d}{d\rho}\right) + 
\frac{{\hbar^2\ell ^2}}{2m^*}\frac{1}{\rho^2} + \frac{1}{8} m^* 
\omega_{c}^2(t) \rho^2 + \frac{1}{2} \frac{\ell \hbar \omega_c}{2},
\end{equation}
and $H_1(t)=\frac{1}{8}m^*\left(2B_{dc}B_{ac}f(t) 
+ B_{ac}^2f^2(t)\right)\rho^2$. 
Here $H_{dc}$ is the standard static Hamiltonian for a charge in a 
homogeneous constant perpendicular magnetic field, that includes the 
para- and dia-magnetic
contributions, with $\omega_c = \frac{e B_{dc}}{m^* c}$. 
With the additional dropping of a term of the form 
$L_z B_{ac}f(t)$ which can trivially be removed by a unitary 
transformation, 
$H_1(t)$ gives the time-dependent contribution to $H$. Note
that $H_1(t)=H_1(t+T_0)$. In the limit in which $H_1(t)$ is much
smaller than $H_{dc}$ 
one can study the modification to the solutions associated to 
$H_{dc}$ by standard time-dependent perturbation theory. As can be 
seen from the
classical phase diagram given in Fig. 1 the boundary between 
regular and chaotic behavior
in fact occurs for $\epsilon=B_{ac}/B_{dc}>1$ and ${\tilde{\omega_c}}>1$. 
We are then led to approximate $H_1(t)$ by,
\begin{equation}
\label{eq:bb}
H_1(t)=\frac{1}{8}m^*\left({(\epsilon\omega_{c})}^2
\sum _{n=-\infty}^{\infty}\delta(t-nT_0)\right)\rho^2.
\end{equation}
With this simplification, the Hamiltonian (\ref{eq:a}) is then 
approximated by the  sum of Eqs. (\ref{eq:b}) and (\ref{eq:bb}).
This choice is also motivated by the following factors: 1. 
Calculational ease: 
the delta function is the paradigm for time-dependent systems 
because one can proceed 
further in the analysis without recourse to drastic approximations; 
2. Effects of chaos: since our primary objective is to explore the 
quantum manifestations of classical chaos, we are more interested 
in the general issues of chaos, rather than specific functional 
forms. Even for a more 
`physical' choice of $f(t)=A\cos(\omega t)$, one can easily show 
that the resulting functional form of $\Omega^2(t)$ can be 
approximated sensibly as above; and 3. Classical 
considerations: as shown in the Appendix, starting from the 
Lorenz force plus Maxwell's equations, one can write the classical 
equations of motion {\it exactly} including the self-induced fields, even for the magnetic field given 
by $B(t) = B_{dc} + B_{ac} T_0\sum_{n=-\infty}^{\infty} 
\delta (t-nT_0)$. Classically, the associated Lagrangian is linear in the 
vector potential.  There are regularization 
problems, however, when using  this form in the quantum Hamiltonian, 
since in this 
case there is an ill defined $A_{ac}^2(t)$ term present. However, the 
model $H=H_{dc}+H_1$ is well defined. 

In order to more clearly see what the relevant parameters in the 
problem 
are, we go over to dimensionless units, defined by rescaling all 
lengths to 
the disk radius $R_0$, all masses by the effective mass $m^*$ and all 
times by the period of the $a.c.$ field, 
$T_0$. Thus, we define 
\begin{mathletters}
\label{allf}
\begin{equation}
\label{eq:f1}
r = \rho/R_0; \qquad 0\le r \le 1,
\end{equation}
\begin{equation}
\label{eq:f2}
\tau = t/T_0 \equiv \frac{\omega_0}{2\pi}t, \qquad
{\tilde\omega_c} = \omega_c/\omega_0, \qquad
{\tilde\hbar} = \frac{\hbar}{m^*\omega_0 R_0^2}.
\end{equation}
\end{mathletters}
In these units, equations (\ref{eq:b}) and (\ref{eq:bb}) become 
\begin{mathletters}
\label{allg}
\begin{equation}
\label{eq:g0}
\tilde H = \tilde H_{dc} + \tilde H_1(\tau)
\end{equation}
\begin{equation}
\label{eq:g1}
\tilde H_{dc} = -\frac{{\tilde\hbar}^2}{2}\left( \frac{d^2}{dr^2} 
+ \frac{1}{r}\frac{d}{dr}
\right) + \frac{\ell^2 {\tilde\hbar}^2}{2 r^2} + \frac{1}{2}
\left(\frac{{\tilde\omega_c}}{2}\right)^2 r^2 + \ell\, {\tilde\hbar} 
\frac{{\tilde\omega_c}}{2} ,
\end{equation}
\begin{equation}
\label{eq:g2}
\tilde H_1(\tau) = \frac{1}{2}\ \eta\ r^2 
\sum_{n=-\infty}^{\infty} \delta (\tau-n), \quad {\rm where}
\end{equation}
\begin{equation}
\eta = \left(\frac{\epsilon\,{\tilde\omega_c}}{2}\right)^2,
\end{equation}
\end{mathletters}
and the corresponding solutions to the time-independent part, 
along with the 
boundary and normalization conditions, are given by
\begin{mathletters}
\label{allh}
\begin{equation}
\label{eq:h1}
\tilde H_{dc}\, \tilde\Psi_{n\ell}(r,\phi) = \tilde E_{n\ell}\, 
\tilde\psi_{n\ell}(r)  \frac{e^{i \ell \phi}}{\sqrt{2 \pi}},
\end{equation}
\begin{equation}
\label{eq:h2}
\tilde\Psi(r,\phi) = \sum_{n=1}^{\infty} \sum_{\ell=-\infty}^{\infty} 
\tilde\psi_{n\ell}(r) \frac{e^{i \ell \phi}}{\sqrt{2 \pi}},
\end{equation}
\begin{equation}
\label{eq:h3}
\tilde\psi_{n\ell}(r = 1) = 0, \qquad {\rm and} \qquad 
\int_{0}^{1} \tilde\psi_{n\ell}^{2}(r) \, r \,dr = 1.
\end{equation}
\end{mathletters}
As was first pointed out by Dingle\cite{dingle}, the physically 
acceptable 
solutions to equations (\ref{allh}) are the Whittaker functions of 
the first 
kind\cite{abramowitz},
\begin{equation}
\label{eq:i}
\tilde\psi_{n\ell}(r) = \sqrt{\frac{2}{N_{n\ell}}}\; 
\frac{1}{r}\;M_{\chi_{n\ell},{\mid \ell\mid}/2}(2\pi F\, r^2),
\end{equation}
where the frustration parameter $F = \frac{\Phi}{\Phi_{0}}$ is the ratio 
of the flux threading the dot to the quantum of flux $\Phi_0=h/2e$. 
The quantities $\chi_{n\ell}$ are related to the eigenvalues via 
\begin{equation}
\label{eq:k}
\chi_{n\ell} = \frac{1}{2}( \tilde E_{n\ell} - \ell ),
\end{equation}
and are determined precisely by the requirement that the wavefunction 
vanishes at the boundary, equation (\ref{eq:h3}), 
$M_{\chi_{n\ell},{\mid \ell\mid}/2}(F) = 0$. 
In the limit of no confinement, $R_0 \rightarrow \infty$, we 
recover the 
usual Laguerre polynomial solutions for the $\tilde\psi_{n\ell}$'s.

The frustration parameter $F$ can also be written as 
\begin{equation}
\label{eq:m}
F = \frac{1}{4\pi}\left(\frac{R_0}{\ell_H}\right)^2, \qquad 
{\rm where} 
\qquad \ell_H=\left(\frac{\hbar c}{eB_{dc}}\right)^{1/2},
\end{equation}
that is, it's proportional to the square of the ratio of the 
confinement 
radius to the magnetic length. When $2\pi F \ll 1$, the problem is 
equivalent 
to that of a nearly free electron, bound by a very weak magnetic 
field, and 
so is amenable to a perturbative treatment. In the opposite limit, 
the 
boundary can essentially be neglected, and we recover the results 
of Dingle 
mentioned previously. It is in the intermediate regime, when the 
two lengths are comparable, that we expect the effects of 
confinement to be 
nontrivial, especially in the presence of strong time-dependent 
fields. 

In principle, we 
are able to cover the entire range of parameter values within the 
same 
framework by means of a numerical evaluation of the Whittaker 
functions. 
However, the Whittaker functions are not very well suited to large 
scale 
computations, because of the time required to evaluate each 
individual 
function. We choose instead to perform most of our calculations in 
a Fourier 
sine basis, which is numerically much faster, and use the Whittaker 
basis as 
a check on our results. Choosing the (orthonormalized) basis 
functions as,
\begin{equation}
\label{eq:eq1}
\chi_{n\ell} = \sqrt{\frac{2}{r}} \sin(n\pi r)  
\frac{e^{i\ell\phi}}{\sqrt{2\pi}},
\end{equation}
one can show, after a straightforward calculation, that the matrix 
elements of 
$\tilde H_{dc}$ are given by,
\begin{eqnarray}
\label{eq:eq2}
(\tilde H_{dc})_{mn} &=& \Bigg\{ \frac{{\tilde\hbar}^2}{2}(n\pi)^2 + 
n\pi (\ell^2-\frac{1}{4}) 
{\tilde\hbar}^2{\rm Si}(2n\pi) + \frac{1}{2}\left
(\frac{{\tilde\omega_c}}{2}\right)^2 
\left(\frac{1}{3}-\frac{1}{2n^2\pi^2}\right) + \ell{\tilde\hbar}
\frac{{\tilde\omega_c}}{2} 
\Bigg\} ~ \delta_{mn} \nonumber \\
&+& \Bigg\{ \frac{\pi}{2}(\ell^2-\frac{1}{4}){\tilde\hbar}^2\left
\{ (m+n){\rm Si}
\left[(m+n)\pi\right] - (m-n){\rm Si}\left[(m-n)\pi\right]\right\} 
\Bigg . \nonumber \\
&+& \Bigg . \frac{1}{2}\left(\frac{{\tilde\omega_c}}{2}\right)^2 
\frac{(-)^{m+n}}{\pi^2} \frac{8mn}{(m^2-n^2)^2} \Bigg\}~ 
(1-\delta_{mn})
\end{eqnarray}
where ${\rm Si}(x)$ is the Sine integral. 
One can similarly compute matrix elements of other needed operators. 

Having worked out a suitable set of basis functions, we now proceed 
to tackle 
the full time-dependent problem. 
The Schr\"odinger equation for the time evolution operator is,
\begin{equation}
\label{eq:n}
i {\tilde\hbar}\,\frac{\partial}{\partial\tau}\,U(\tau,\tau_0) = 
( \tilde H_{dc} + 
\tilde H_1(\tau) )\,
U(\tau,\tau_0).
\end{equation}
Since we have a periodic system, $\tilde H(\tau+1) = \tilde 
H(\tau)$, 
from the Floquet theorem\cite{casati2}, it is sufficient to 
determine the 
one-period time evolution operator $U(\tau_0 + 1,\tau_0)$, from 
\begin{mathletters}
\label{allo}
\begin{equation}
\label{eq:o1}
i {\tilde\hbar}\,\frac{\partial}{\partial\tau}\,U(\tau) = ( \tilde 
H_{dc} + \tilde V_1(\tau) )\,U(\tau) \, , \quad 0 < \tau \le 1,
\end{equation}
\begin{equation}
\label{eq:o2}
\tilde V_1(\tau) = \tilde V \, \delta(\tau-1), \qquad {\rm where} 
\qquad 
\tilde V = \frac{1}{2}\ \eta\ r^2,
\end{equation}
\end{mathletters}
where the parameter $\eta$ has been defined previously. 
All the information about the dynamics of the system is contained 
within this 
Floquet operator, since $\Phi(r,\phi,\tau+1) = U\,
\Phi(r,\phi,\tau)$, where 
$\Phi$ is the total wave function. Because of the  
periodic $\delta$-kicked dynamics, we can immediately integrate 
equation 
(\ref{eq:o1}) to get 
\begin{equation}
\label{eq:p}
U_\ell(1,0) = \exp\left(-\frac{i}{{\tilde\hbar}}\,\tilde V\right)\,
\exp\left(-\frac{i}{{\tilde\hbar}}\,\tilde H_{dc}\right).
\end{equation}
The subscript $\ell$ has been attached to $U$ to emphasize that the 
evolution operator has been restricted to that single $\ell$ value. In 
other 
words, states with different values of $\ell$ evolve independently, 
an immediate 
consequence of the conservation of angular momentum in this system. 
The 
rightmost exponential operator in equation (\ref{eq:p}) evolves the 
wave function from just after the `kick' at $\tau=0$ to just before 
the kick at one period  
under the influence of $\tilde H_{dc}$, while the operator to it's 
left 
propagates it from just before to just after the kick at a period. 

Since $U$ is an Unitary operator, the spectrum of it's eigenvalues 
can be represented as 
\begin{equation}
\label{eq:q}
U_\ell\,\phi_{n\ell} = e^{i\,\varepsilon_{n\ell}}\,\phi_{n\ell}.
\end{equation}
The set of eigenvalues $\{\varepsilon_{n\ell} \in (0,2\pi]\}$, are 
collectively 
known as the  Quasi-energy Eigenvalues (QEE), and the eigenfunctions 
$\{\phi_{n\ell}\}$ as the Quasi-energy Eigenfunctions (QEF) of $U$. 
The investigation of the quantum dynamics of the system is completely 
equivalent to determining the nature of the QEE and QEF. The 
fundamental task 
is thus to obtain the Quasi-energy Spectrum (QES) of the evolution 
operator given by equation (\ref{eq:p}).

\section{Classical Dynamics}

We begin the discussion of the behavior of the model by looking at 
it's 
classical dynamics. The classical Hamiltonian corresponding to the 
quantum 
one given by equations (\ref{allg}) is, 
\begin{mathletters}
\label{alls}
\begin{equation}
\label{eq:s0}
\tilde H = \tilde H_{dc} + \tilde H_1(\tau)
\end{equation}
\begin{equation}
\label{eq:s1}
\tilde H_{dc} = \frac{1}{2} p_r^2
+ \frac{J^2}{2 r^2} + \frac{1}{2}
\left(\frac{{\tilde\omega_c}}{2}\right)^2 r^2 + J 
\frac{{\tilde\omega_c}}{2} ,
\end{equation}
\begin{equation}
\label{eq:s2}
\tilde H_1(\tau) = \frac{1}{2}\ \eta\ r^2 
\sum_{n=-\infty}^{\infty} \delta (\tau-n),
\end{equation}
\end{mathletters}
where $p_r$ is the radial momentum and $J$ is the {\it conserved} 
angular 
momentum. To make quantitative correspondences between the classical 
and quantum results, we always set the numerical values of the 
angular 
momenta in the two cases to be equal, {\sl i.e.}, we set $J=\ell\,
{\tilde\hbar}$. 

In between the `kicks' at a period, and as long as it does not hit 
the boundary 
at $r=1$, the electron's motion is governed by the 
static Hamiltonian $\tilde H_{dc}$. The equation of motion in this 
case is  
\begin{equation}
\ddot r = -\left(\frac{{\tilde\omega_c}}{2}\right)^2 r + 
\frac{J^2}{r^3},
\end{equation}
whose solution, in terms of the energy $E$, 
\begin{equation}
\label{eq:t}
E = \frac{1}{2} p_r^2
+ \frac{J^2}{2 r^2} + \frac{1}{2}
\left(\frac{{\tilde\omega_c}}{2}\right)^2 r^2 + J 
\frac{{\tilde\omega_c}}{2},
\end{equation}
is given by 
\begin{equation}
\label{eq:u}
\left( \begin{array}{c} 
r(\tau) \\
p_r(\tau) 
\end{array} \right) = 
\left( \begin{array}{c}
\sqrt{\frac{2}{{\tilde\omega_c}}\left[ b + a 
\sin\left\{ {\tilde\omega_c}(\tau-\tau_0) 
+ \sin^{-1}\left(\frac{\frac{1}{2}{\tilde\omega_c} r_0^2 - b}{a}
\right)\right\}\right]} \\
\frac{a}{r(\tau)}\cos\left\{ {\tilde\omega_c}(\tau-\tau_0) + \sin^{-1}
\left(\frac{\frac{1}{2}{\tilde\omega_c} r_0^2 - b}{a}\right)\right\}
\end{array} \right),
\end{equation}
where
\begin{equation}
\label{eq:v}
b = 2E/{\tilde\omega_c} - J \quad {\rm and} \quad a = \sqrt{b^2 - 
J^2}.
\end{equation}
Here, $r_0$ and $\tau_0$ are initial conditions. For a given energy 
$E$, the 
motion is constrained by the centrifugal barrier on one side, 
and the smaller 
of the wall radius (equal to 1) and the constraint imposed by the 
attractive quadratic potential on the other:
\begin{mathletters}
\label{allw}
\begin{equation}
\label{eq:w1}
r_{min} \le r(\tau) \le {\rm Min}\{r_{max},1\}, \qquad {\rm where}
\end{equation}
\begin{equation}
\label{eq:w2}
r_{min} = \sqrt{\frac{2}{{\tilde\omega_c}}(b-a)},\qquad {\rm and} 
\qquad
r_{max} = \sqrt{\frac{2}{{\tilde\omega_c}}(b+a)}.
\end{equation}
\end{mathletters}
Note that the equations of motion are nonlinear here, even in the 
walls' 
absence. The effect of collision with the wall (or centrifugal 
barrier) is 
simply to reverse the direction of motion:
\begin{equation}
\label{eq:x}
\left( \begin{array}{c}
r(\tau_c^+) \\
p_r(\tau_c^+)
\end{array} \right) = 
\left( \begin{array}{cc}
1 & 0 \\
0 & -1
\end{array} \right)
\left( \begin{array}{c}
r(\tau_c^-) \\
p_r(\tau_c^-)
\end{array} \right),
\end{equation}
where $\tau_c$ is the time of collision with the wall (or barrier). 
Finally, 
the effects of the kicks at $\tau=n$ are obtained by integrating the 
equations of motion over an infinitesimal duration around $n$:
\begin{equation}
\label{eq:y}
\left( \begin{array}{c}
r(n^+) \\
p_r(n^+)
\end{array} \right) = 
\left( \begin{array}{cc}
1 & 0 \\
\eta r & 1 
\end{array} \right)
\left( \begin{array}{c}
r(n^-) \\
p_r(n^-)
\end{array} \right).
\end{equation}
If we denote  the mapping 
due to 
the `free' evolution of the particle under the influence of  
$H_{dc}$ by $M_0$ 
(equations (\ref{eq:u})), that due to the walls by  equation 
by $M_{wall}$ (equations (\ref{eq:x}), 
and the mapping due to the kick by $M_{kick}$ 
(equations (\ref{eq:y})), 
then the complete one-period map is typically given by 
the 
product of several $M$'s for a given energy, {\sl i.e.},
\begin{equation}
M_{T} = \left( M_0\cdot M_{wall}\right)^N \cdot M_{kick}.
\end{equation}
In general, the map is very complicated, and very sensitive to 
initial 
conditions. By recording the values at each successive period, 
we obtain 
a surface-of-section of the trajectory of the particle in phase space.

There are three independent parameters in the problem: 
${\tilde\omega_c}$, $\epsilon$ and 
${\tilde\hbar}$. However, for quantitative correspondences to be 
made later with the 
quantum results, as mentioned earlier, we keep the angular momentum 
$J=\ell{\tilde\hbar}$ 
fixed, which reduces the number of parameters to the first two. 
The transition 
to chaos is manifested in the parameter space spanned by 
$(\epsilon,{\tilde\omega_c})$ 
(see Fig. 1). All 
of our subsequent results refer to this space. We did investigate 
the effects 
of varying $J$ by varying ${\tilde\hbar}$ for fixed $\ell$, and 
the results are even 
quantitatively very similar.

The first (and most obvious) evidence of chaotic behavior is seen 
in the Poincar\'e surface of section in $(r,p_r)$. In Figures 
2(a)--(d) 
we show the sections corresponding to $\epsilon$ values of 0.5, 
1.5, 1.95 and 
2.5, respectively, while ${\tilde\hbar}=0.01$, 
${\tilde\omega_c}=2.0$ and $\ell=5$ are held fixed. 
(The reason for this particular choice has to do with the 
$(\epsilon,{\tilde\omega_c})$ `phase diagram' for 
this system, which is explained in more detail shortly.)  
In the quasi-integrable regime (Figs. 2(a),(b)), 
the phase space is dominated by invariant tori, which are close to 
those of the unperturbed problem. As the value of ${\tilde\omega_c}$ 
is increased, 
the tori begin to break up, and isolated chaotic islands begin 
to appear (Fig. 2(c)), 
until finally, all evidence of invariant curves disappears and 
all we see 
is the uniform chaotic sea (Fig. 2(d)). These values of  
$(\epsilon,{\tilde\omega_c})$ 
corresponding to the integrable, intermediate and chaotic regions 
will 
be retained throughout what follows to make comparisons between the 
classical and quantum results.  

Corresponding to the transition from regular to chaotic behavior, 
we begin to see the appearance of diffusive 
growth in the averaged energy (or squared momentum) of a localized 
ensemble of initial conditions. Figure 3(a) shows the average 
energy as a 
function of time for the parameters corresponding to the 
quasi-integrable regime, while Figure 3(b) corresponds to parameter 
values in the chaotic regime. 
In contrast to the behavior in the quasi-integrable regime, 
where the energy $E$ is regular, oscillatory quasiperiodic 
functioning of time around a 
constant value, in the chaotic regime $E$ grows linearly 
(or $p_r$ grows quadratically) with time. 
(Here and subsequently, `time' refers to stroboscopic time, 
just after every kick ). 

A quantitative measure of the degree of chaos in the system is to 
calculate the largest Lyapunov exponent. (In our reduced two-dimensional 
phase space since the flow is Hamiltonian, the Lyapunov exponents come in pairs 
of opposite sign.) Because our phase-space is bounded, we use a 
slightly modified approach from that used for an unbounded system to the calculation 
of the exponent, as outlined in Reichl\cite{reichl}. 
The (largest) Lyapunov exponent is defined by,
\begin{equation}
\label{eq:eq17}
\lambda_n(\tau,{\rm{\bf X}}_{0,0},{\rm{\bf Y}}_{0,0}) = 
\frac{1}{n\tau}
\sum_{j=1}^{n} \ln\left(\frac{d_j}{d_0}\right),
\end{equation}
where $d_0 = |{\rm{\bf Y}}_{0,0}~-~{\rm{\bf X}}_{0,0}|$ is the 
Euclidean distance between the position of neighboring 
trajectories labelled by 
{\bf X}$_{0,0}$ and {\bf Y}$_{0,0}$, and $\{d_j\}, j=1,\dots,n$ are the 
sequence of distances generated between the trajectories at $n$ 
successive time steps. If $d_0$ is not too big, then the limit, 
$lim_{n\uparrow\infty} 
\lambda_n(\tau,{\rm{\bf X}}_{0,0},{\rm{\bf Y}}_{0,0}) = 
\lambda({\rm{\bf X}}_{0,0})$ exists, and 
is independent of both $d_0$ and $\tau$. Furthermore, 
$\lambda({\rm{\bf X}}_{0,0})$ is zero if 
{\bf X}$_{0,0}$ is chosen in a regular region, while it is positive 
if {\bf X}$_{0,0}$ is chosen to lie in a chaotic region. 

With the help of the Lyapunov exponent we 
constructed the `phase diagram' shown in Fig.1 for this system 
in the $(\epsilon,{\tilde\omega_c})$ parameter space in the 
following fashion. For a given 
set of parameters $(\epsilon,{\tilde\omega_c})$, we choose a very 
large number (typically 
$10^6$) initial conditions ${\rm{\bf X}}_{0,0}$ 
spread uniformly in $(r,p_r)$ phase space. Next we randomly choose a nearby
phase space point ${\bf Y}_{0,0}$ within a circle of radious $d_0$,
centered about ${\bf X}_{0,0}$. We calculate the Lyapunov exponent,
using formula (25), from the successive evaluation of the distances
$d_j$ for each $j$ iteration of the mapping. This process is repeated
for several nearby ${\bf Y}_{0,0}$ trajectories.
When the Lyapunov exponent reaches saturation 
we average the resulting value over the set of initial conditions to
find $\lambda$.
If this asymptotic value is positive, the system is defined as 
chaotic. To put a 
stricter criterion on the degree of chaos, we choose a threshold 
value of 
the exponent $\lambda_c$ beyond which the system is in the regime of 
hard chaos. We set $\lambda_c$ arbitrarily to the value 1, but as a 
check 
we generated Poincar\'e phase portraits to confirm chaos by looking 
for 
featureless ({\sl i.e.,} no invariant tori) phase portraits. In 
this way, 
by varying the parameters $(\epsilon,{\tilde\omega_c})$ in a 
continuous fashion over the 
whole plane, running the map repeatedly and obtaining the resulting 
$\lambda$'s, we obtained the `phase diagram' for this system, 
including 
a distinct `phase boundary' separating the quasi-integrable and 
hard chaos 
regions. Of course, this phase boundary depends on the precise 
value of 
the cutoff $\lambda_c$ we choose. Nevertheless, we checked that on 
varying 
the cutoff $\lambda_c$, the phase boundary shifts only slightly and 
furthermore, the shape of the boundary remains qualitatively the 
same. 
Indeed, to a high degree of precision, the phase boundary can be 
fitted by 
\begin{equation}
\label{eq:eq18}
{\tilde\omega_c} = C(\lambda_c)/\epsilon,
\end{equation}
where $C(\lambda_c)$ is a constant which depends on the value of 
the cutoff. 
Figure 1 shows the phase diagram for a cutoff $\lambda_c=1$.

%

%%%%%%%%%%%%%%%%%%%%%%%%%%%%%%%%%%%%%%%%%%%%%%%%%%%%%%%%%%%%%%%%%%%%%%%%%%%%

We observe from the classical Poincar\'e sections that there is a 
symmetry 
line in the $(r,p_r)$ plane. This arises from the time-reversal 
invariance present in the problem as follows. Consider a particle 
kicked at $\tau=0$. 
The position $r_0$ remains unchanged, while the momentum changes : 
$p_r^{(+)} = p_r^{(-)} + \eta\ r_0$. Denoting $p_r^{+}$ by $p_0$, 
then at 
time $0^{(-)}$ the particle had momentum $p_r^{(-)} = p_0 -\eta\ 
r_0$. Taking 
into account the fact that the angular momentum is conserved, we 
see that 
propagating a particle {\it forward} in time from $(r_0,p_0)$ is 
the same 
as propagating it {\it backward} from $(r_0,\eta\ r_0-p_0)$. Thus, 
the motion 
is symmetric about the line $p_r = -\frac{1}{2}\eta\ r$. This 
symmetry is, 
of course, present in the quantum problem also, where it will be 
exploited when 
calculating the Husimi distributions of the QEF's. In the 
classical case, 
we exploit its existence to plot the stable manifolds around 
hyperbolic fixed 
points, which 
are otherwise very difficult to do because 
of their extreme sensitivity to perturbations.

Although the map is very complicated, there are a few periodic orbit 
cases that  one can analytically study. By following the trajectory 
of the periodic orbit in phase space, and given the mapping 
equations, 
we can reconstruct the initial conditions giving rise to the orbit. 
For example, 
the fixed point shown in the Fig.5 
(for $\ell=5$, ${\tilde\hbar}=0.008$, ${\tilde\omega_c}
=2\sqrt{2}$ and $\epsilon=1.0$), labeled F, 
is given by $r_0 = 0.75528003154206\ldots$,
$p_0 = 2.43838534012017\ldots$.

\section{Quantum to Classical correspondence}
\label{subsubsec:sqfad}

As mentioned in the introduction, one of the clear quantum 
manifestations 
of classical chaos (QMCC) 
emerges when one compares the spectral properties  of specific 
model systems 
as appropriate parameters are tuned to classically produce a 
transition 
from integrable to completely chaotic regimes. In this section we 
follow 
the general thinking developed in Random Matrix Theories (RMT) to 
implement 
different tests to quantify the spectral properties of the model.
These properties are obtained from  a direct diagonalization 
of the one-period time evolution matrix. For the results presented 
here
we vary the value of $\epsilon$ while keeping $J$, ${\tilde\hbar}$ 
and ${\tilde\omega_c}$ fixed, so 
as to go from the integrable to the chaotic regime in the phase 
diagram that coincide with the values considered in the classical 
case. We note that the appropriate RMT statistical ensemble is a COE
rather than a CUE, because this model has a false-T breaking symmetry.

Next we discuss the RMT tests and their application to the results 
obtained for the QEE of our quantum dot model.

\subsection{Nearest neighbor QEE distributions}

A local measure often used in RMT is the distribution of 
nearest-neighbor
energy level separations, $P(s)$, where 
$s=\varepsilon_{n+1}-\varepsilon_n$. 
In the extreme integrable and chaotic regimes it has been 
established\cite{bohigas2,jose} that $P(s)$ takes the Poisson or 
Wigner distribution forms,  
\begin{equation}
\label{eq:qee}
P_P(s) = e^{-s} \qquad {\rm and} \qquad
P_W(s) =  {\pi\over 2}\,s\,e^{-{\pi\over 4}s^2},
\end{equation}
respectively.
A convenient and often successful parameterization of the $P(s)$ 
obtained in the transition between $P_P$ to $P_W$ is provided
by the Brody interpolation formula\cite{brody}:
\begin{equation}
P_{\nu}(s) = \gamma(\nu+1)\,s^{\nu}\,\exp(-\gamma s^{\nu+1}),
\end{equation}
where $\gamma = \left[\Gamma\left({\nu+2\over\nu+1}\right)\right]^
{\nu+1}$, and $\Gamma(x)$ is the Gamma function. This distribution is 
normalized and, by construction, has mean spacing 
$\langle{s}\rangle=1$. 
We recover the Poisson case taking $\nu =0$ and Wigner for $\nu =1$.
A criticism to the Brody distribution is, however, that there is 
no first 
principles  justification for its validity. The fact remains that it 
does fit the specific results found when considering explicit 
model systems. Results of the transition, as parameterized 
by $\nu$, are shown in Figure 5.

We also calculated higher-order eigenvalue spectral 
correlations\cite{bohigas1}. 
The average number of levels in an interval of length L is
$<n(L)> = {1\over L}\sum_{\alpha< }n(\alpha ,L),$
where the $< >$ stands for spectral average, and $n(\alpha ,L)$ 
is the number of
levels in an interval of length $L$ starting at $\alpha$ and 
ending at $\alpha +
L$. Also important are the various moments of the level 
distribution. The one
considered here is the second moment of the average number of 
levels in 
a given stretch of length $L$ of the spectrum, the $\Sigma^2(L)$ 
statistic 
\begin{equation}
<\Sigma ^{2}(L)> = \left <(n(\alpha ,L) - <n(\alpha ,L)>)^2 \right >.
\end{equation}
Another often calculated statistic is the Dyson-Mehta 
$\Delta _3(L)$  which measures the stiffness of the  spectrum.  
This is defined by
\begin{equation}
{\Delta _3(L,\alpha )}={1\over L}{min_{A,B}}
{\int _{\alpha }^{\alpha +L}}
{[{\tilde N}(x)-Ax-B]^2}\,dx,
\end{equation}
where $\tilde N(x)$ is the unfolded number density. In our case 
there is no 
need to unfold the spectrum since it is fully contained between 0 and 
2$\pi$;   
$\Delta _3$ is just the least mean square 
deviation of ${\tilde N}(x)$ from the mean  straight line
behavior. This statistic is directly proportional to the
$<{\Sigma ^2}>$ by $\Delta _3 (L)={2\over {L^4}}
\int _0^L(L^3-sL^2x+x^3)
\Sigma ^2(x)dx$, and thus can be calculated for the Circular 
Orthogonal Ensembles (COE) 
as well\cite{mehta}. The specific theoretical predictions for the 
averaged 
${<{\Delta_3(L)}>}={1\over L}{\sum _{\alpha }
\Delta_3(L,\alpha )}$, are ${{\Delta_3^{(COE)}(L)}}\>={{1\over \pi 
^2}}\ell n\{L\} -0.007,\> \>$ and
${<{\Delta_3^{(Poisson)}(L)}>}\>={L\over 15}\> $.
These results are correct in the asymptotic limit valid for 
$15\leq L$.

In Fig.6 we present our results for $<\Delta_3>$ and $<\Sigma ^2>$.
In these figures one clearly sees the transition 
from Poisson-like (dashes) to COE-like (solid line) behavior 
as $\epsilon/{{\tilde\hbar}}$ is varied. We note that the 
$\Delta_3$ statistic does not saturate in the COE limit, even 
for the maximum 
interval $L$  that we looked at, as would be expected from 
semiclassical 
arguments originally proposed by Berry\cite{berry}. Furthermore, 
note that for the largest $L$ 
considered the Poisson limit does not present the knee seen in 
other completely
integrable systems as was found before\cite{jose}. 
All in all the results shown in Fig.6 are consistent
with what we have come to expect for the transition
between regular and chaotic regions.

\subsection{ Quasi-energy eigenfunction statistics}

Here we consider the statistical properties of the eigenfunction
overlaps  with the natural basis vectors.
it has been conjectured  \cite{alhassid} that as
the classical motion changes from chaotic to regular, this 
distribution of overlaps
can be represented by a $\chi^2$-distribution in $\nu$ degrees of 
freedom, with  $\nu$ varying  from 1  in the chaotic regime
(the Porter-Thomas limit) to 0 in the regular region (the Poisson limit):
\begin{equation}
P_{\nu}(y) = {(\nu /2)^{\nu /2}\over \Gamma (\nu /2)}
\enskip y^{\nu /2 -1}\enskip \exp(-\nu y/2).
\end{equation}
Here $y \equiv \mid \langle\lambda |nl\rangle\mid ^2$, 
where $\mid{\lambda}\rangle$ 
label the QEF and 
$|{nl}\rangle$ label a set of $N$ orthogonal basis vectors. 
(The $y$'s have been  rescaled so that $\langle{y}\rangle=1$.) We 
have tested this hypothesis for 
the overlap strengths for the same parameter values as for the 
quasi-energy eigenvalue statistics. 
The results are shown in Fig. 7, plotted on a logarithmic scale. 
These results show the general trend of decreasing $\nu$ as we 
cross the phase boundary from regular to chaotic classical motion. 
However, we note that as we go from the COE to the Poisson 
limits, the fits to the
$\chi^2$ get worse. Note especially the shift of the maxima away 
from zero.  
This discrepancy is connected  to the fact that the results are strongly
basis 
dependent when not in the universal COE limit.

\subsection{Semiclassical correspondences}

We can now make a direct comparison between the classical and quantum 
results by employing a phase space approach. To do this, we
use the Husimi representation of the QEF. The
Husimi distribution, interpreted as a probability density, is a
coarse-grained version of the Wigner function which goes smoothly 
to the
semiclassical limit\cite{chang}. In practice, the most often used 
technique of
coarse-graining is to take the overlap of the QEF with 
coherent oscillator
states. For the radial coordinate the coherent state is
\begin{equation}
\Psi_{r_0,p_0}^G(r) = ({\sigma\over\pi{\tilde\hbar}})^{1\over 4}\,
\exp\left\{-{\sigma\over 2{\tilde\hbar}}(r - r_0)^2 + i{p_0\over 
{\tilde\hbar}}(r - {r_0\over 2}) \right\},
\end{equation}
which is a minimum-uncertainty Gaussian wavepacket
centered at $(r_0,p_0)$, with root mean-squared deviations given by
$\Delta\rho = \sqrt{{\tilde\hbar}/2\sigma}$, $\Delta p = 
\sqrt{{\tilde\hbar}\sigma/2}$,
and $\sigma\,$ is the `squeezing' parameter. This parameter is 
adjusted
when making comparisons to the classical phase-space plots. The 
Husimi distribution of a single QEF $\phi_{\varepsilon}(r)$, 
is then defined by
\begin{equation}
{\cal F}_{\phi_{\varepsilon}}(r_0,p_0) = \left| \>\int_0^1
\Psi_{r_0,p_0}^G(r)\,\phi_{\varepsilon}(r)\,dr\> \right|^2.
\end{equation}
The Husimi distribution is obtained by scanning through the values of
$(r_0,p_0)$ in the region of interest in phase space, and the 
result is
compared with the classical surface-of-section.
We begin the comparison by noting the symmetry about the line 
$p=-\eta r$
in the Husimi contour plots in Fig.8. As mentioned earlier, this
feature carries over from the classical results for the same 
reasons as there,
and it is in fact used to effectively halve the numerical effort.

All calculations reported here were carried out for relative 
cyclotron  
frequency ${\tilde\omega_c}=2\sqrt{2}$, angular momentum quantum 
number $\ell=5$, relative 
{\it a.c.} to {d.c.} field strength $\epsilon=1$ and scaled 
${\tilde\hbar}=0.008$. In this case, all terms in the Hamiltonian 
are comparable in magnitude, which means that we are in a
non-perturbative regime. Furthermore, we can clearly see both 
from the phase 
diagram and the surface-of-section that this places the system on the 
order-chaos border, where the dynamics is quite `mixed'. 
A few calculations were done for different values of the 
parameters, but no 
new qualitative features emerged. In choosing the value of
${\tilde\hbar}$, we were guided by the following considerations. 
The value of
${\tilde\hbar}$ has to be small enough so that the system is well 
into the
semiclassical regime, yet large enough so that the dimension of 
the truncated
Hilbert space $N$ (which grows as the inverse square of 
${\tilde\hbar}$) is large enough
to preserve unitarity. Moreover, $N$ has to be such that the largest 
eigenenergy of $H_{dc}$ has to be larger than the maximum energy 
of the classical
particle in the region of interest in phase space. All the
interesting features seen in this model are manifested in this regime.
Finally, the classical conserved angular momentum $J$ was kept 
identical to the quantum value, $\ell{\tilde\hbar}$.

The classical analysis was carried out for different values of the 
angular momentum $J$\cite{badri1}. First, we iterated
a single (arbitrarily chosen) initial condition several thousand 
times,
which typically leads to the chaotic background as shown in the 
figures.
Embedded in this background are KAM tori centered around elliptic 
fixed points,
defined by choosing appropriate initial conditions. In Figure 8, we 
show
several such tori, and in particular, a fixed point of period 4 
which was determined earlier analytically. Also shown
in each of the figures is a hyperbolic fixed point of order 6, 
marked by
its stable and unstable manifolds. The fixed points were determined 
by
using a modified Powell method of determining zeros of coupled 
nonlinear
sets of equations\cite{recipes}. This method, like all 
multidimensional root-finding
techniques, requires a good initial guess to converge to the fixed 
point, but
once given it determines the root and the Monodromy matrix (the 
Jacobian
or the determinant of the linearized version of the map equations 
reliably 
and accurately. The fixed points are elliptic, parabolic or 
hyperbolic if the
discriminant obtained from the eigenvalues ({\it i.e.}, 
$(Trace)^2-4\cdot (Determinant)$) is negative, zero or positive, 
respectively.
In all cases, it was verified, within numerical error, that the map 
was
area-preserving, {\it i.e.}, the determinant was equal to one. The 
unstable
manifold was obtained by iterating the map along the direction 
given by
the eigenvector corresponding to the eigenvalue larger than one. The
{\it stable} manifold is given by the time reversed version of the 
unstable one.

Comparison of the Husimi distributions 
${\cal F}_{\phi{\varepsilon}}(r_0,p_0)$ 
with
the classical phase space plots show some striking similarities. 
There are, for many QEF, many structures which unmistakably 
correspond to elliptic, parabolic and hyperbolic periodic orbits, as 
seen in Fig. 8. For example, the Husimi representation of one 
of the QEF sits 
on top of the analytic period-two fixed point marked as F.  Also, 
seen in  the figure are  Husimis which peak {\it   exactly} on top
of the unstable hyperbolic period-6 fixed point, referred to in the
literature as `scars'\cite{jensen}. This correspondence is so 
robust, in fact, that
often when a good guess to the {\it classical} hyperbolic fixed 
points
are unavailable, the Husimis are used as a guide to the location 
of the
fixed point (being unstable, hyperbolic fixed points cannot be 
located without
a very good initial guess). These enhanced probability densities are
conjectured to play as important a role in quantum mechanics as the
hyperbolic orbits play in classical chaos. Finally, a rare but 
persistent
occurrence in all the cases considered is that of a single Husimi 
distributions
peaked simultaneously over {\it both} elliptic and hyperbolic fixed 
points,
reflecting a purely quantum-mechanical tunneling across the KAM tori.
Here we have only shown representative results of the correspondence 
between Husimi distributions and classical solutions.

\section{Experimental Feasibility}
\label{sec:exp}

Before concluding, we present some experimental scenarios where the 
predicted effects may be observable.

A `typical' GaAs-AlGaAs semiconductor quantum dot device 
\cite{marcus1},\cite{levy} has a radius $R_0$ of between 0.1 and 10$\mu$m, 
a sheet density $n\sim 10^{11}$ cm$^{-2}$, and a mobility 
$\mu\sim 2.65 \times 10^5$ cm$^2$/V$\cdot$s. The typical level spacing  
$\Delta\epsilon \sim 0.05$ meV or $\sim 500$ mK. The operating temperatures 
is generally of the order of 0.1 K, so $kT\sim 0.01$ meV is smaller than 
$\Delta\epsilon$, and thus the discrete spectrum can be accessed. 
A typical elastic mean free 
path $\l_\phi\sim 10\mu$m, and the phase coherence length varies between 
15 and 50 $\mu$m. The power injected is typically $<$ 1 nW, which avoids the 
problem of electron heating.

Given these parameters, we can estimate in physical units the field 
strengths and frequencies required to observe the effects predicted by 
our model. Let us first calculate these assuming a dot radius $R_0\sim 1\mu$m.  
The fundamental kick frequency $\omega_0$ in our problem can be deduced from 
Eqs. \ref{eq:f2} as
$\omega_0 = \hbar/(m^*R_0^2\tilde\hbar)
         \simeq [1/\tilde\hbar]\,2\times 10^{9} {\rm s}^{-1} .$
From this, we can deduce the required $d.c.$ and $a.c.$ magnetic field 
magnitudes:
\begin{eqnarray}
\label{allexp-2}
B_{dc} &=& \frac{\omega_0 m^* c}{e^*}\,{{\tilde\omega_c}} \nonumber \\
       &\simeq& 20 \frac{{\tilde\omega_c}}{{\tilde\hbar}} \,{\rm Gauss} \\
B_{ac} &=& \epsilon B_{dc} \simeq 
           20 \frac{\epsilon{\tilde\omega_c}}{{\tilde\hbar}} \,{\rm Gauss} .
\end{eqnarray}
Finally, the Larmor frequency associated with  the $a.c.$ field is given by 
$\omega_{ac} = \epsilon {\tilde\omega_c} \simeq 
\epsilon{\tilde\omega_c}/(\tilde\hbar)\, 
2\times 10^7 \,{\rm s}^{-1}.$
The dot radius $R_0$ in Ref. \cite{levy} is about 5$\mu$m. For this 
radius, the frequency and $d.c.$ magnetic field magnitudes are,
$\omega_0 \simeq \,8\times 10^{7} {\rm s}^{-1}{{\tilde\hbar}}^{-1}$ and 
$B_{dc} \simeq 0.8 \tilde\omega_c/\tilde\hbar\,{\rm Gauss}. $

With these values, we can see what physical parameters correspond to the 
integrable and chaotic regimes. We fix ${\tilde\hbar}=0.1$, and choose as  
representative parameters $(\epsilon,{\tilde\omega_c})^{(reg)} = (0.1,0.1)$
where the motion is regular, and the parameters 
$(\epsilon,{\tilde\omega_c})^{(chaos)} = (2.0,2.0)$ where the motion is 
chaotic. Then, for $R_0\sim 1\mu$m, the frequency and $a.c.$ fields 
corresponding to the regular regime are, 
\begin{equation}
\omega_0^{(reg)} \simeq 20 \,{\rm GHz}, \,\,\,\, 
B_{ac}^{(reg)} \simeq 20 \,{\rm Gauss} ,
\end{equation}
while those corresponding to the chaotic regime are, 
\begin{equation}
\label{allexp-6}
\omega_0^{(chaos)} \simeq 20 \,{\rm GHz} ,\,\,\,\, 
B_{ac}^{(chaos)} \simeq 800 \,{\rm Gauss} .
\end{equation}
For the case $R_0 \sim 5\mu$m case, the frequencies and fields are, for the 
regular regime, 
\begin{equation}
\label{allexp-7}
\omega_0^{(reg)} \simeq 800 \,{\rm MHz}, \,\,\,\, 
B_{ac}^{(reg)} \simeq 0.08 \,{\rm Gauss} ,
\end{equation}
and for the chaotic regime, 
\begin{equation}
\label{allexp-8}
\omega_0^{(chaos)} \simeq 800 \,{\rm MHz}, \,\,\,\, 
B_{ac}^{(chaos)} \simeq 32 \,{\rm Gauss}. 
\end{equation}

With the appropriate techniques of measurement, for example by using an 
array of $\sim 10^5$ {\it isolated} quantum dots to increase the magnitude of the 
signal, and using a highly sensitive electromagnetic superconducting 
microresonator to measure the response, as was done by Reulet, {\sl et. al.} 
in Ref. \cite{reulet} to measure the dynamic conductance of mesoscopic rings 
threaded by  Aharonov-Bohm fluxes. We believe that an experimental realization 
of this system is feasible. 
\section{Conclusions}
\label{sec:conc}

We have shown that the model of an electron in a rigid quantum dot 
structure subject to constant and periodically kicked 
magnetic fields shows a transition to chaos, depending on 
the relationship between the strengths of the fields and 
the cyclotron frequency of the steady field. This relationship is 
characterized by a `phase diagram' in parameter space shown in 
Fig. 1. The nature of various periodic orbits were investigated. 
The quantum signatures of this transition are evidenced in two 
measures. First, as the classical system goes from 
integrable to chaotic, the statics of the quasienergy 
spectrum follow the route from Poisson-like to COE-like. Second, 
the contour plots of the Husimi distribution of the 
quasienergy eigenfunctions clearly 
exhibit the phenomenon of `scarring' over unstable periodic orbits.  
Finally, we have presented some experimental ranges of the parameters 
where the effects of chaos in the system may be observable. 
To sum up, all tests applied to the classical quantum 
correspondence are in full agreement with the established 
quantum manifestations of classical chaos.  The many electron 
problem will be treated elsewhere
\cite{badri2}.

%\section*{Acknowledgments}
\acknowledgments
We thank G. Chu for useful discussions.
This work has been supported  by  Office of Naval Research grant
number ONR-N00014-92-1666 and by NSF grant DMR-95-21845.
%
%%%%%%%%%%%%%%%%%%%%%%%%%%%%%% Appendix %%%%%%%%%%%%%%%%%%%%%%%%%%%%%%%
%
\appendix
\section*{}

In this appendix, we show that the classical particle and field 
equations of 
motion can be written exactly for a periodically kicked magnetic 
field. Starting from the Lorenz force equation,
\begin{mathletters}
\label{allapa}
\begin{equation}
\label{eq:apa1}
m^*\frac{d^2{\bf r}}{dt^2} = m^*\frac{d{\bf v}}{dt} = 
e^*\left\{\frac{{\bf v}}
{c}\times {\bf B}(t) + {\bf E}(t) \right\},
\end{equation}
\begin{equation}
\label{eq:apa2}
{\rm where}\qquad{\bf B}(t) = \left( B_{dc} + B_{ac} 
T_0\sum_{n=-\infty}^
{\infty}\delta(t-nT_0) \right){\hat e_z} \equiv 
\left\{ B_{dc} + B_{ac}\Delta(t) \right\}{\hat e_z},
\end{equation}
\begin{equation}
\label{eq:apa3}
{\rm and}\qquad {\bf E}(t) = -\frac{1}{c}\frac{\partial}
{\partial t}{\bf A}(t) 
= \frac{B_{ac}}{2c} {\dot\Delta}(t)\,\left({\bf r}
\times{\hat e_z}\right).
\end{equation}
\end{mathletters}
Then, on substituting Eqs.(\ref{eq:apa2}) and (\ref{eq:apa3}) in 
Eq.(\ref{eq:apa1}), and using the definition of $\omega_c$, we get,
\begin{equation}
\label{eq:apb}
\frac{d{\bf v}}{dt} = \omega_c \left({\bf v}\times{\hat e_z}\right) + 
\epsilon\omega_c \left({\bf v}\times{\hat e_z}\right) \Delta(t) + 
\frac{\epsilon\omega_c}{2}\left({\bf r}\times{\hat e_z}\right)
{\dot\Delta}(t).
\end{equation}
Using the standard property of the delta function, 
$\int{f(x)\delta^{\prime}
(x-a)\,dx} = -f^{\prime}(a)$, the last term becomes
($\epsilon=\frac{B_{ac}}{B_{dc}}$),
\begin{equation}
\label{eq:apc}
\frac{\epsilon\omega_c}{2}\left({\bf v}\times{\hat e_z}\right)
\Delta(t).
\end{equation}
Thus, the {\it exact} equations of motion can be written as,
\begin{equation}
\label{eq:apd}
\frac{d{\bf v}}{dt} = \omega_c \left({\bf v}
\times{\hat e_z}\right)\left\{ 
1 + \frac{\epsilon}{2} T_0\sum_{n=-\infty}^{\infty}
\delta(t-nT_0)\right\}.
\end{equation}
Note that the only difference we have from including the induced 
{\bf E} field is a factor of 1/2 in the kicked component of the 
{\bf B} field. 

The reason the same analysis cannot be done the same 
way in the quantum problem is that there it is the 
vector potential that is the relevant dynamical variable.
Thus if we use an {\bf A} = {\bf A}$_{dc}$ 
+ {\bf A}$_{ac}(t)$ with {\bf A}$_{ac}(t) \simeq 
\sum_{n=-\infty}^{\infty}B_{ac}(\rho)\delta(t-nT_0)$, we see that we have a 
mathematical ambiguity in the definition of {\bf A}$_{ac}^2$. 
Nonetheless, one can carry out the nonrelativistic analysis with  
our model Hamiltonian that contains, we believe, the essential physics 
of the problem and yet is mathematically tractable.

% 
%%%%%%%%%%%%%%%%%%%%%%%%%%%%%% References %%%%%%%%%%%%%%%%%%%%%%%
%

%

%
%%%%%%%%%%%%%%%%%%%%%%% Figure Captions %%%%%%%%%%%%%%%%%%%%%%
%
%
\begin{figure}
\caption{ Classical phase diagram for the problem, obtained from 
the Lyapunov exponent, calculated as explained in the text. 
The shape is fairly insensitive to 
the value of the threshold chosen to characterize hard chaos. 
The circles denote the parameters explicitly 
considered in the classical to quantum comparisons.}
\label{fig1}
\end{figure}

\begin{figure}
\caption{ Poincar\'e surfaces of section in the $(r,p_r)$ plane. 
The values 
of $\epsilon$ are (a) 0.5, (b) 1.5, (c) 1.95 and (d) 2.5. 
${\tilde\hbar}=0.01$, $\ell=10$ 
and ${\tilde\omega_c}=2.0$ are held fixed. We observe a gradual 
breakup of the invariant 
tori until eventually there is no more structure present in the 
phase space.}
\label{fig2}
\end{figure}

\begin{figure}
\caption{ Average energy of an ensemble of points as a function of 
time. In (a) $\epsilon=0.5$, (b) $\epsilon=1.0$, and (c) 
$\epsilon=2.5$, all other parameters as  above. The first 
is stable and oscillatory, the second shows a 
quadratic growth in time (see text), while the third exhibits 
quasi-linear (diffusive) growth, corresponding to particles 
diffusing through the chaotic sea.}
\label{fig3}
\end{figure}

\begin{figure}
\caption{ Schematic trajectory of a period-4 orbit, corresponding 
to ($\epsilon\,$,${\tilde\omega_c}$)=$(1.0,2.828427\ldots)$ .}
\label{fig4}
\end{figure}

\begin{figure}
\caption{ Nearest-neighbor spacing statistic $P(s)$,  
the parameters being the same as in 
Figure 2. Note the gradual movement away from the Poisson to the COE 
distribution, characterized by the Brody parameters $\nu$ 
given by (a) 0.27 (b) 0.52, and (c) 1.0, for the parameters of 
Fig. 2(a),(c) and (d), respectively.}
\label{fig5}
\end{figure}

\begin{figure}
\caption{ (a) $\,<\Delta_3(L)>$ and (b) $\,\Sigma^2(L)$ statistics 
for the same 
parameters as in Figure 2. Again, we see that as $\epsilon$ 
increases, the 
statistics go from being close to Poisson-like (dashed line) to  
COE-like (solid line).}
\label{fig6}
\end{figure}

\begin{figure}
\caption{ (a)-(d) Distribution of amplitude overlaps  of the $QEF$
with the natural basis states for the same parameter values  as in 
Fig.2. 
Close to the COE limit,  (d), the amplitudes are nearly gaussian or 
Porter-Thomas randomly distributed. Away from this limit the 
distributions
are not well fitted by the $\chi^2$ distributions, with a significant 
difference
seen close to the Poisson limit. This discrepancy is explained in 
the text.
The values of $\nu$ from the fits are, (a) $0.14$, (b) $0.27$, 
(c) $0.63$, and (d) $0.9$.}
\label{fig7}
\end{figure}

\begin{figure} 
\caption{ Contour plots of the Husimi distribution of three 
QEF's. The Husimis labelled A correspond
to the period-4 solution, while the one labelled B is another 
example of an enhanced probability distribution over an 
elliptic fixed point of period 4. Finally, we also see a 
Husimi distribution of a QEF which corresponds to
the period-6 hyperbolic orbit marked by its stable 
and unstable manifolds - a `scarred' eigenfunction. 
(The rectangle at the top-right corner indicates the 
uncertainties $\Delta X,\,\Delta P$.}
\label{fig8}
\end{figure}

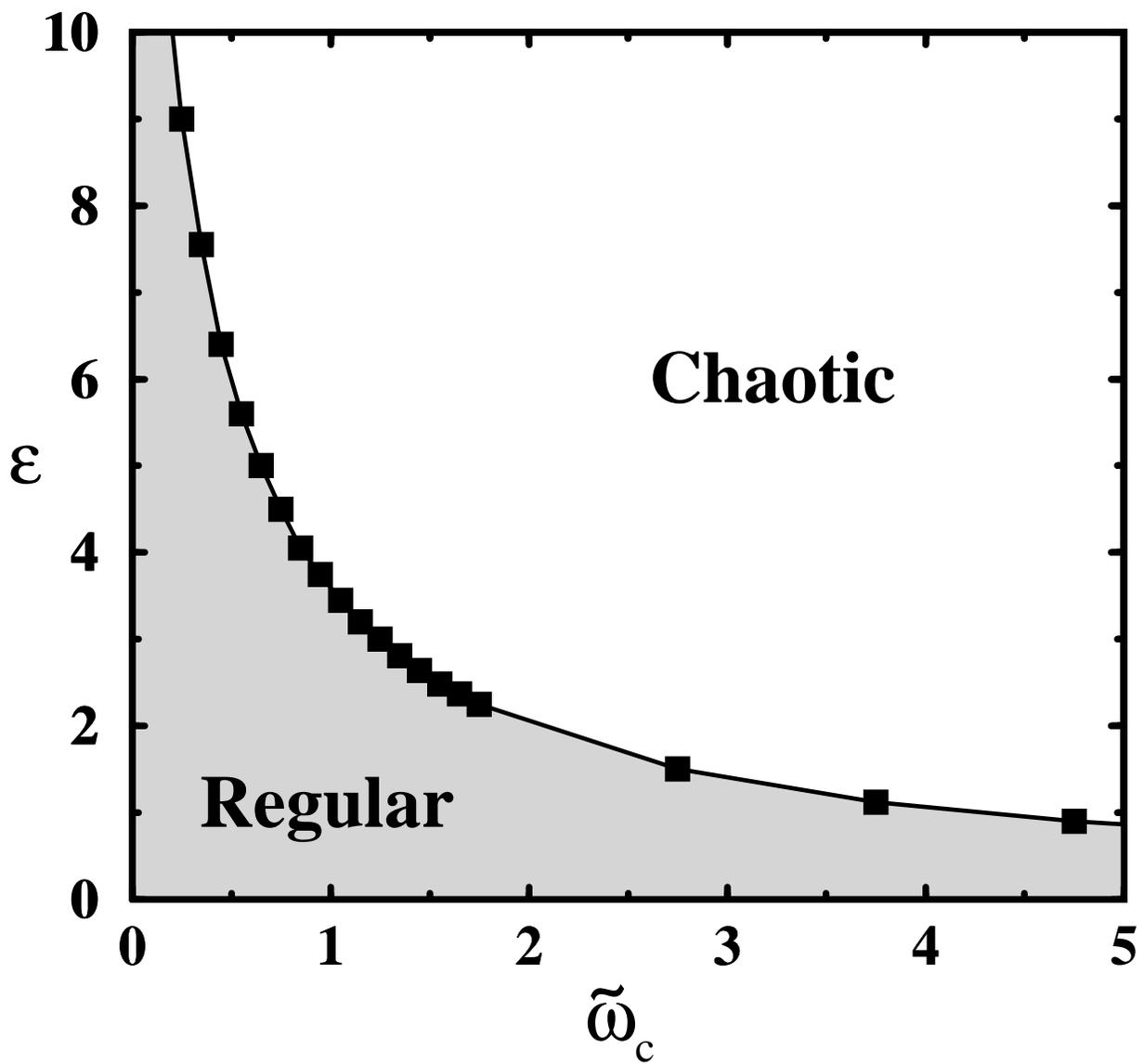
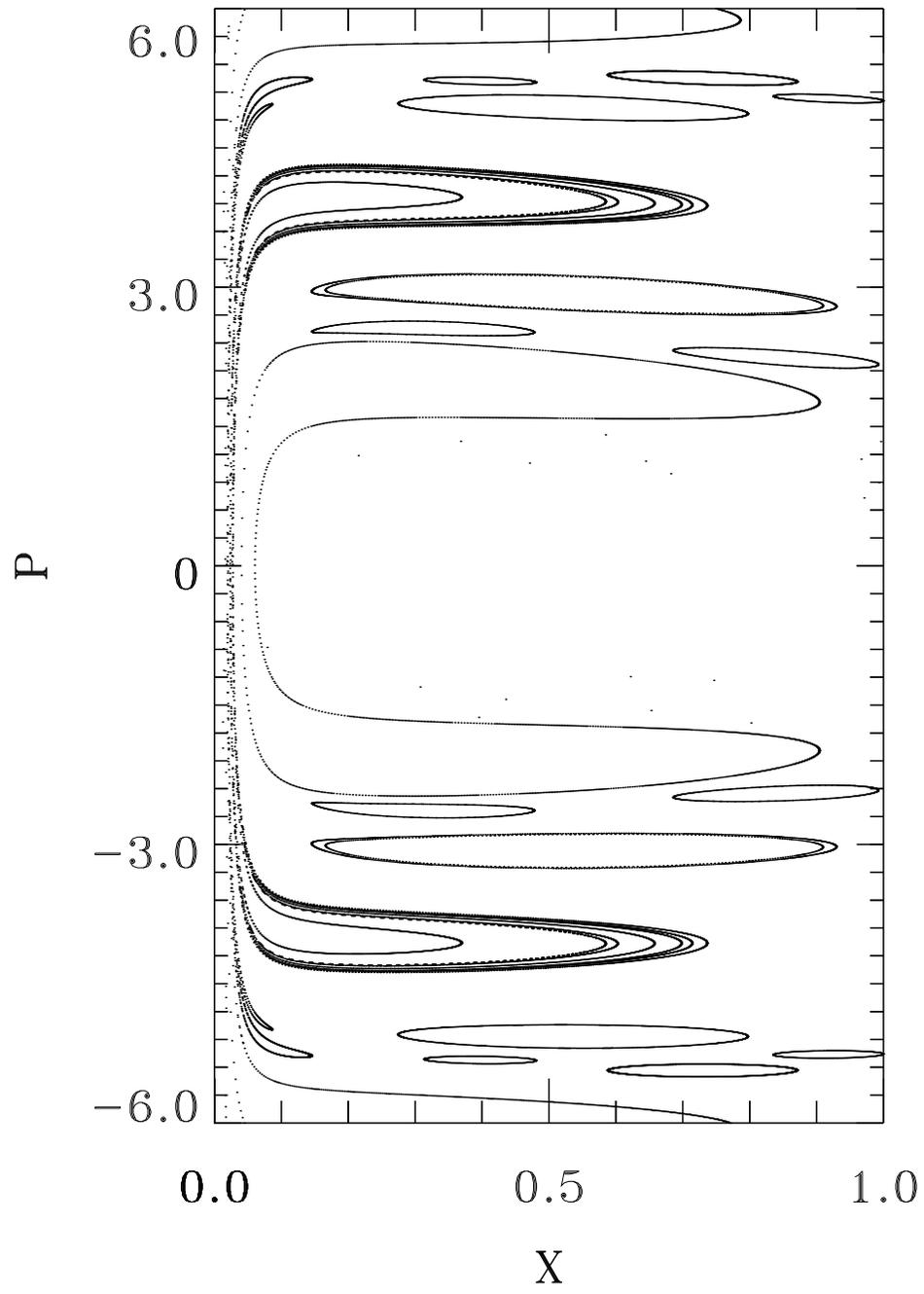
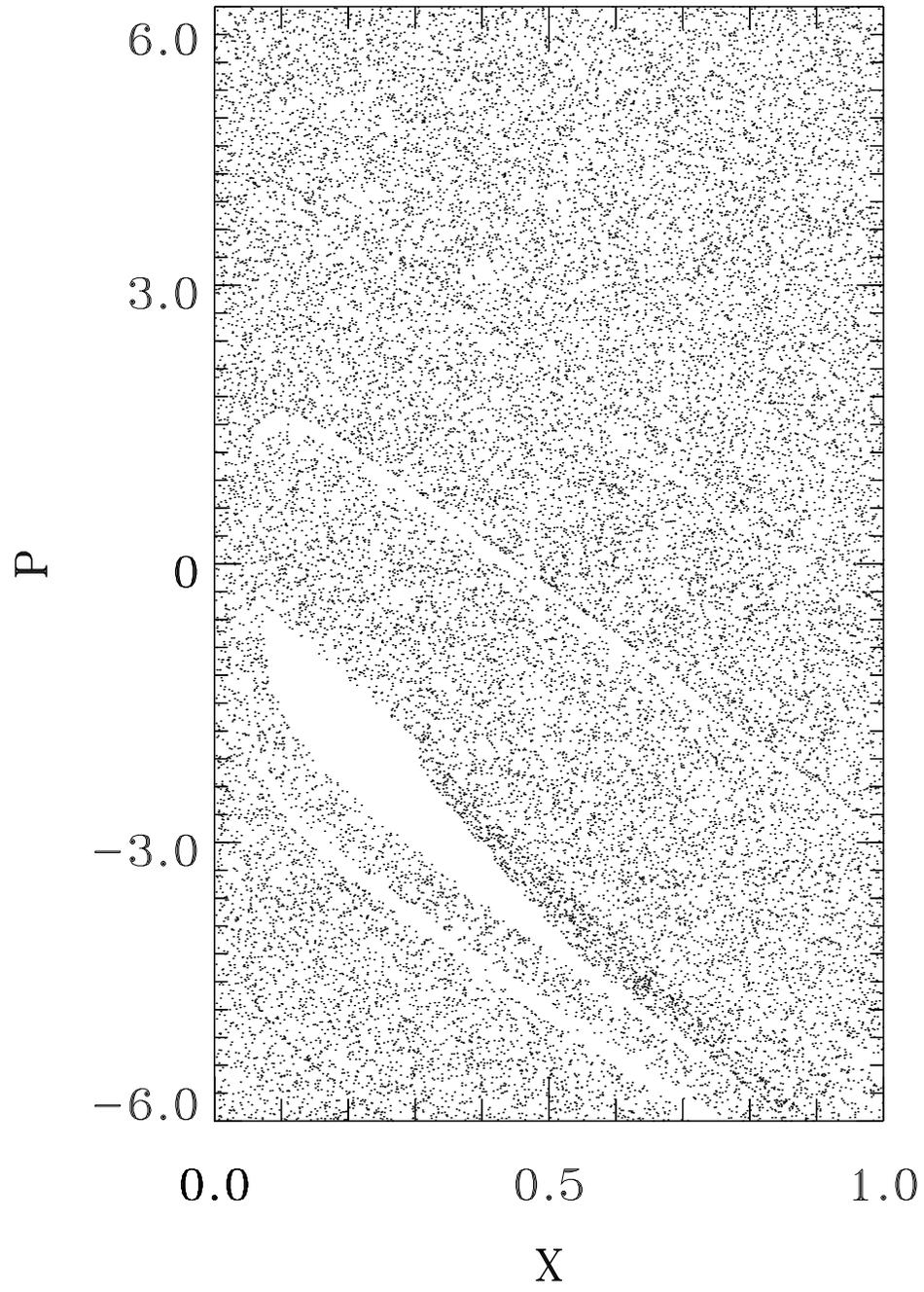
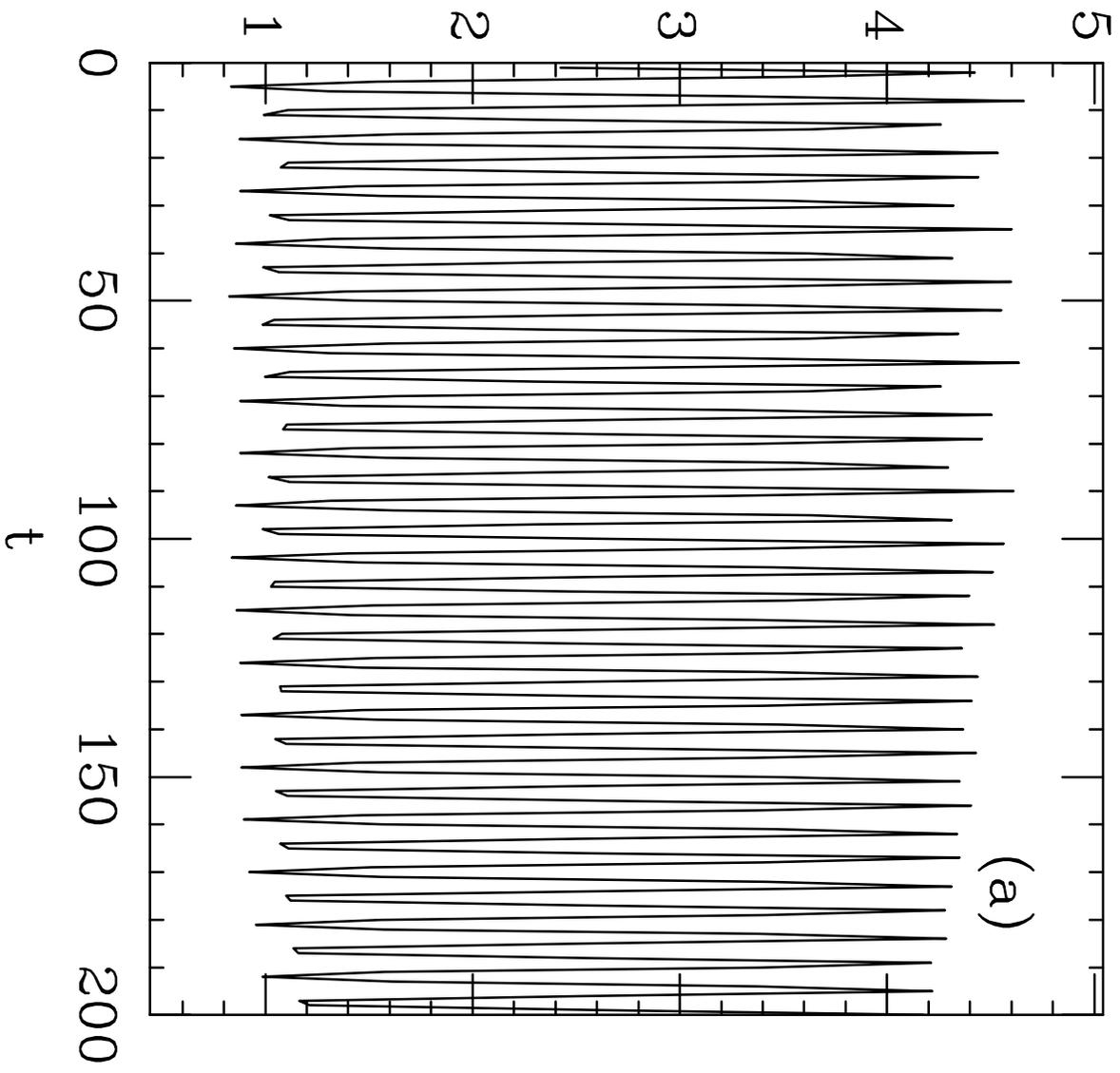
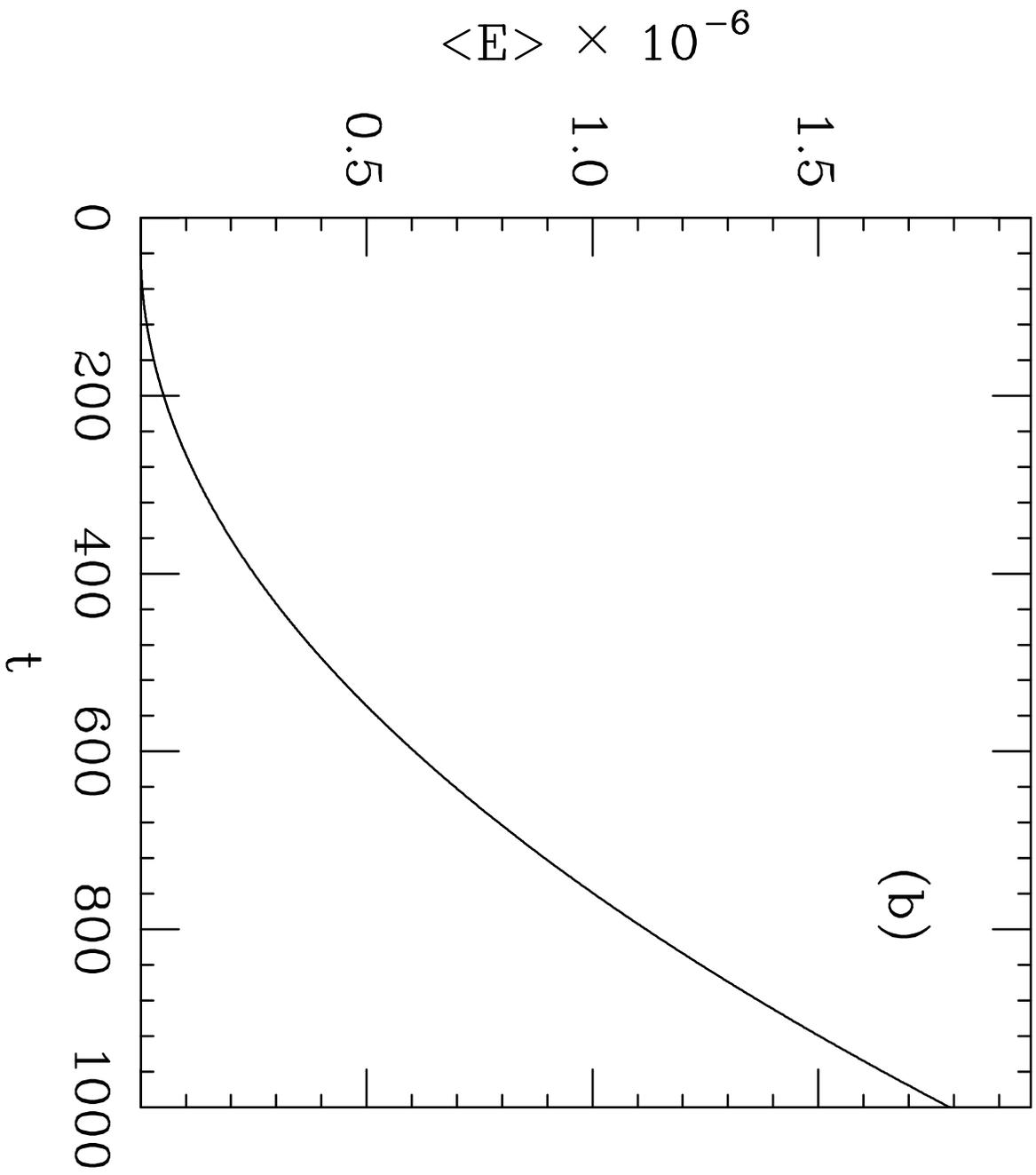
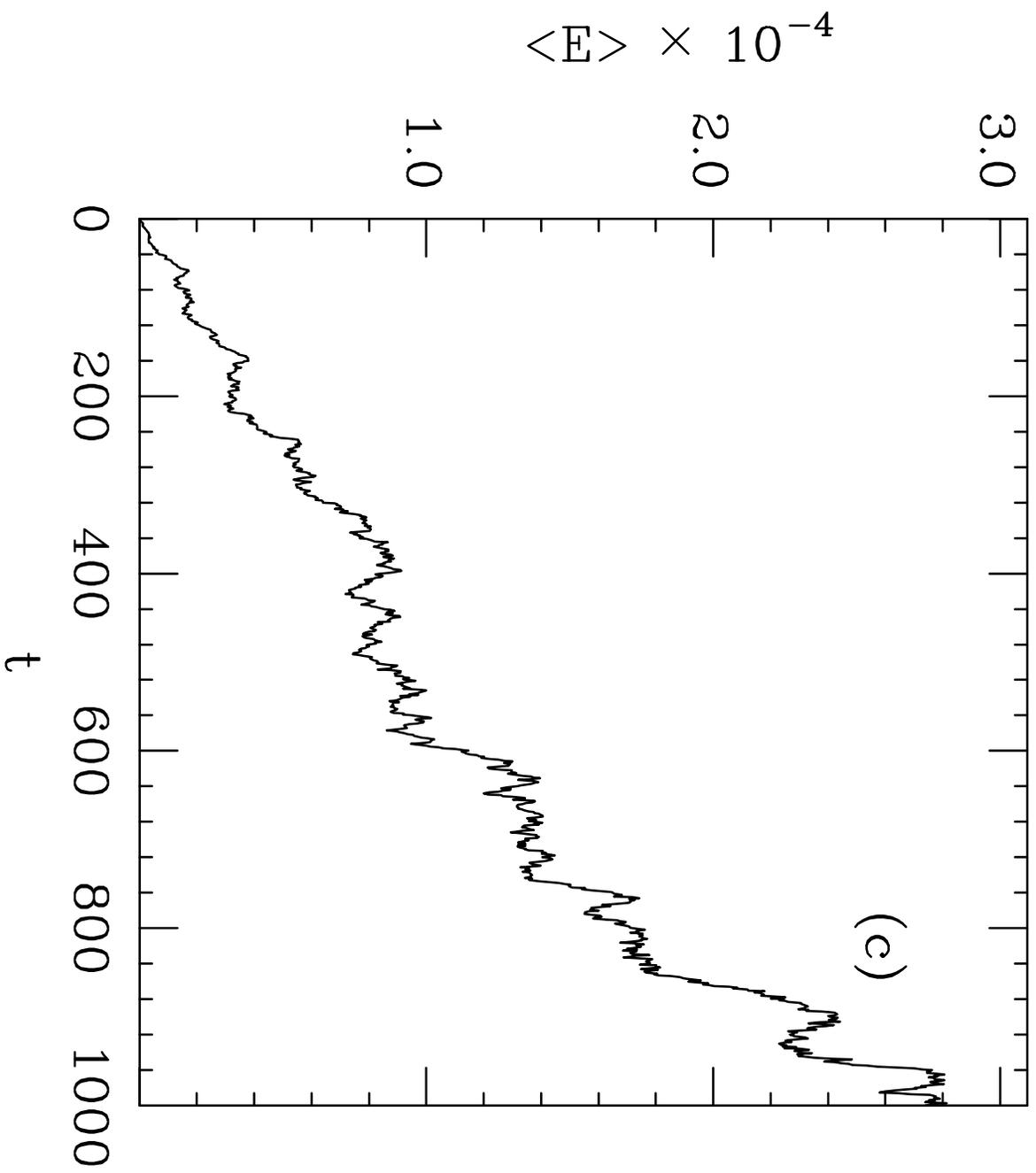
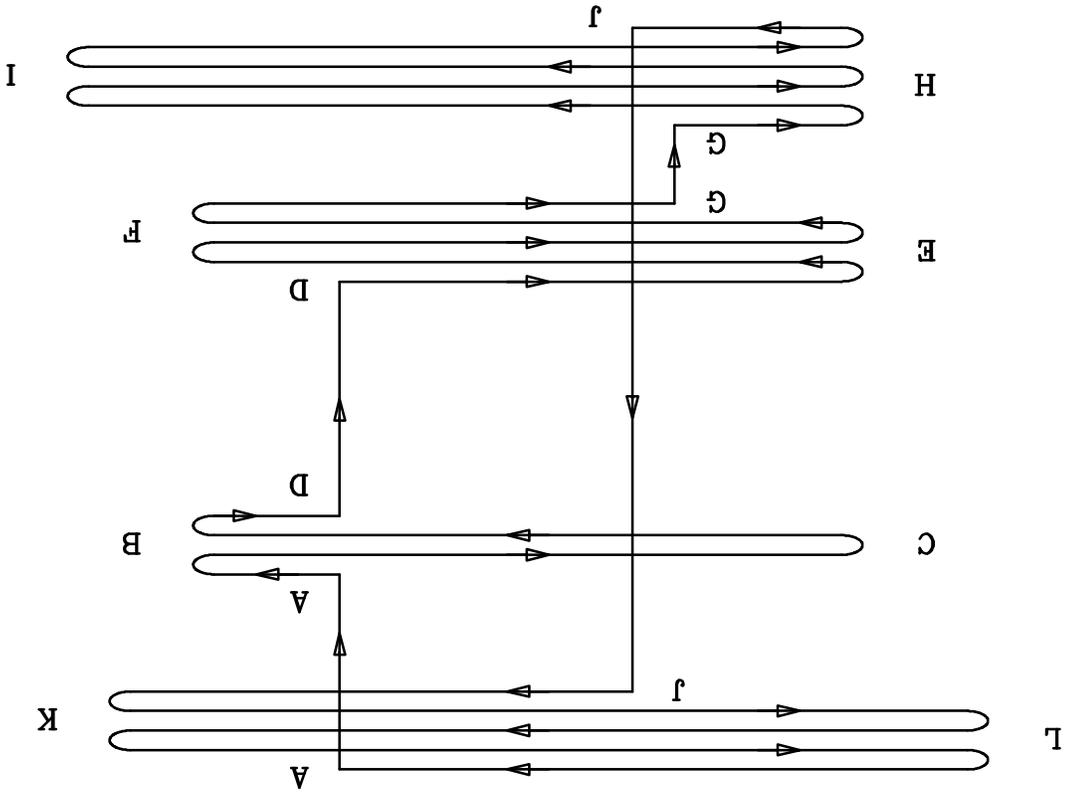
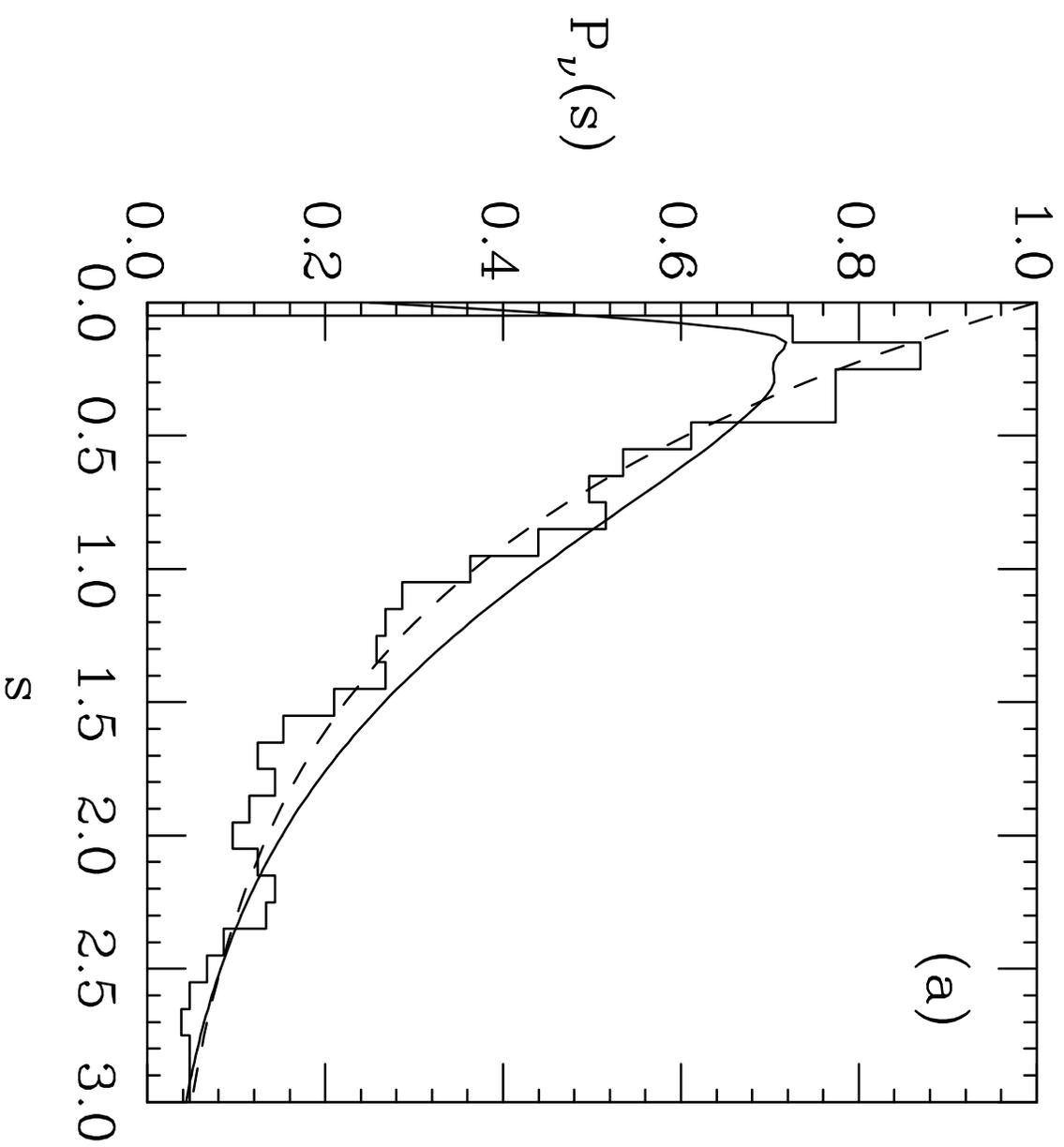
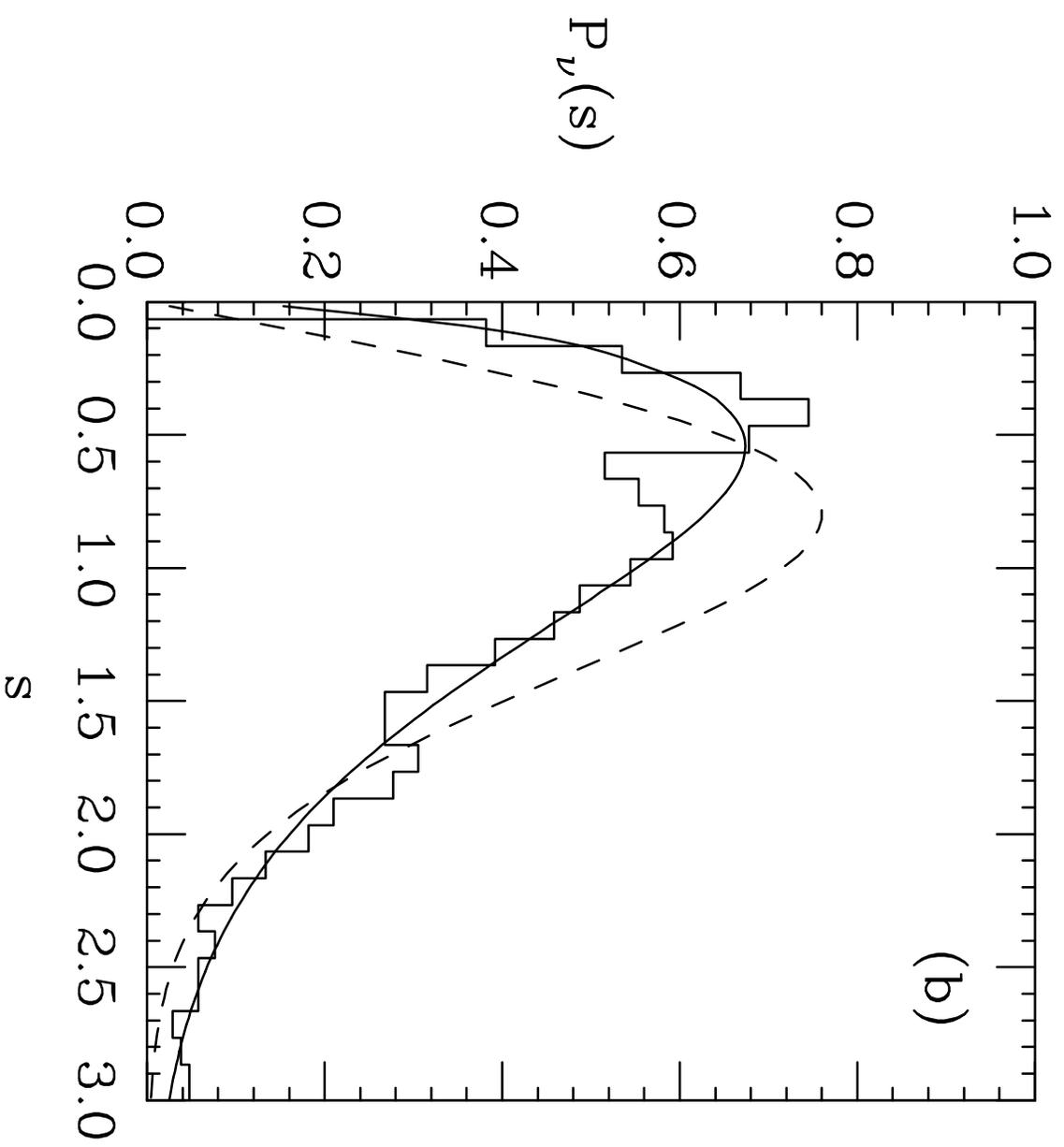
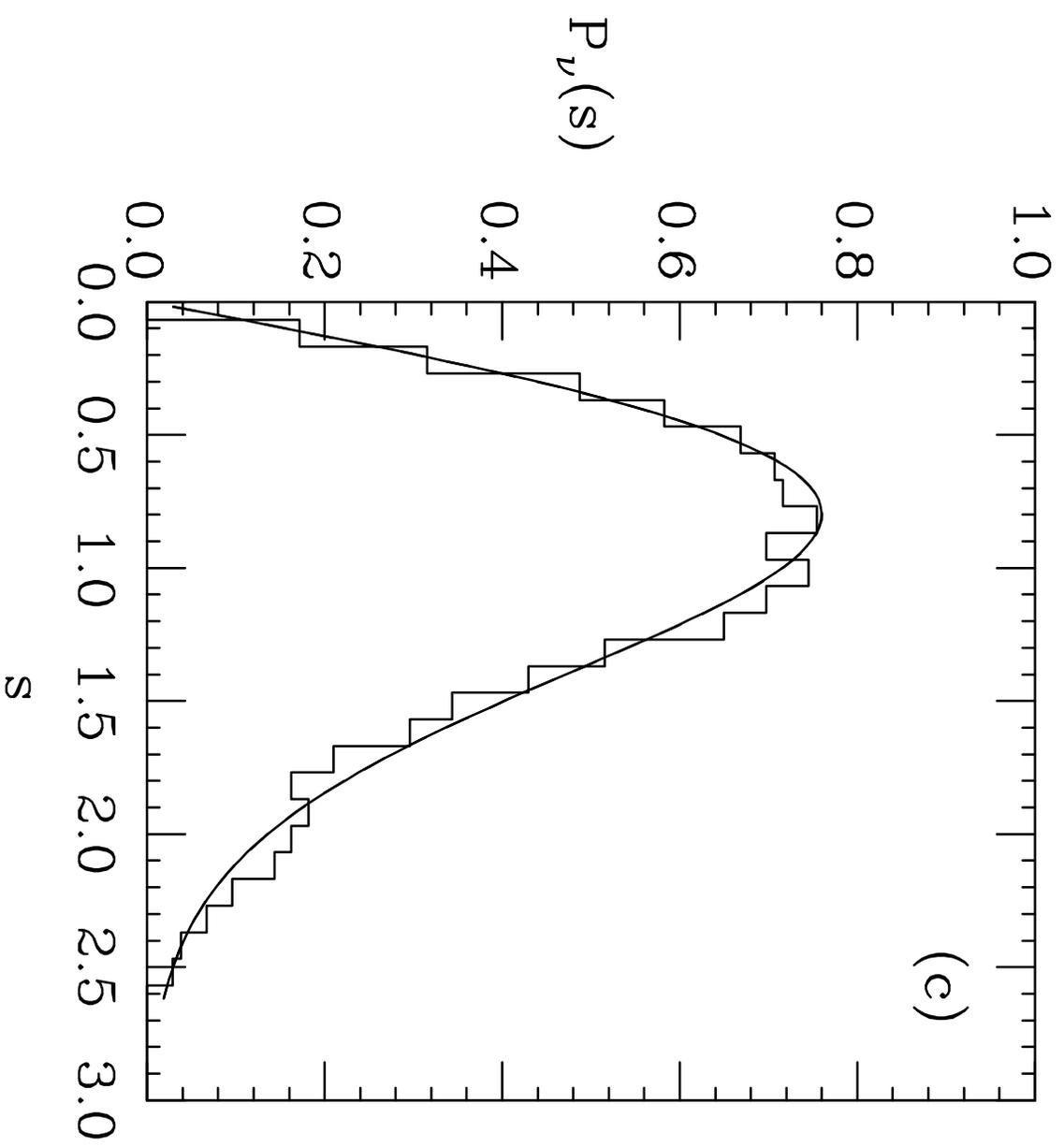
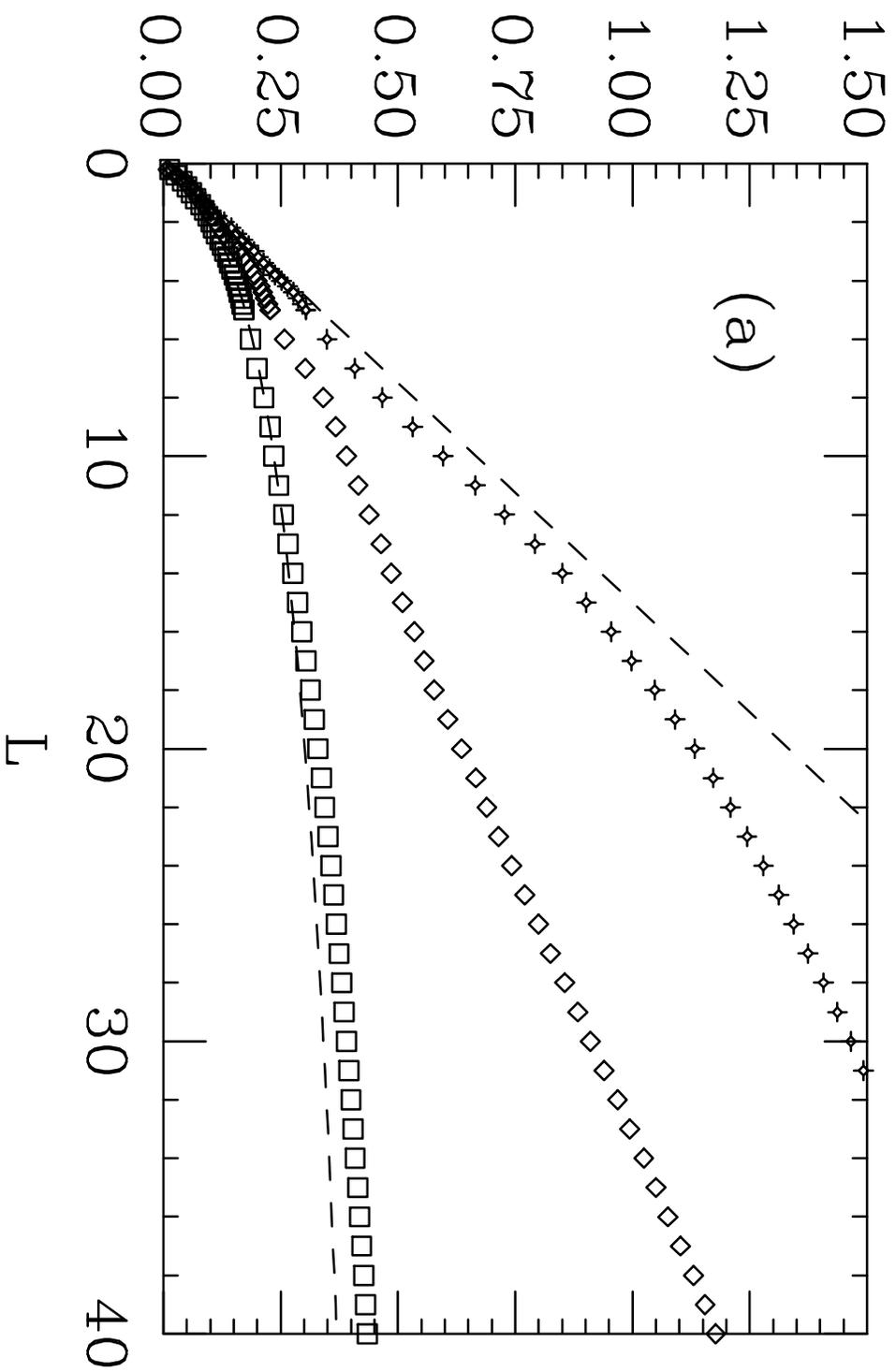
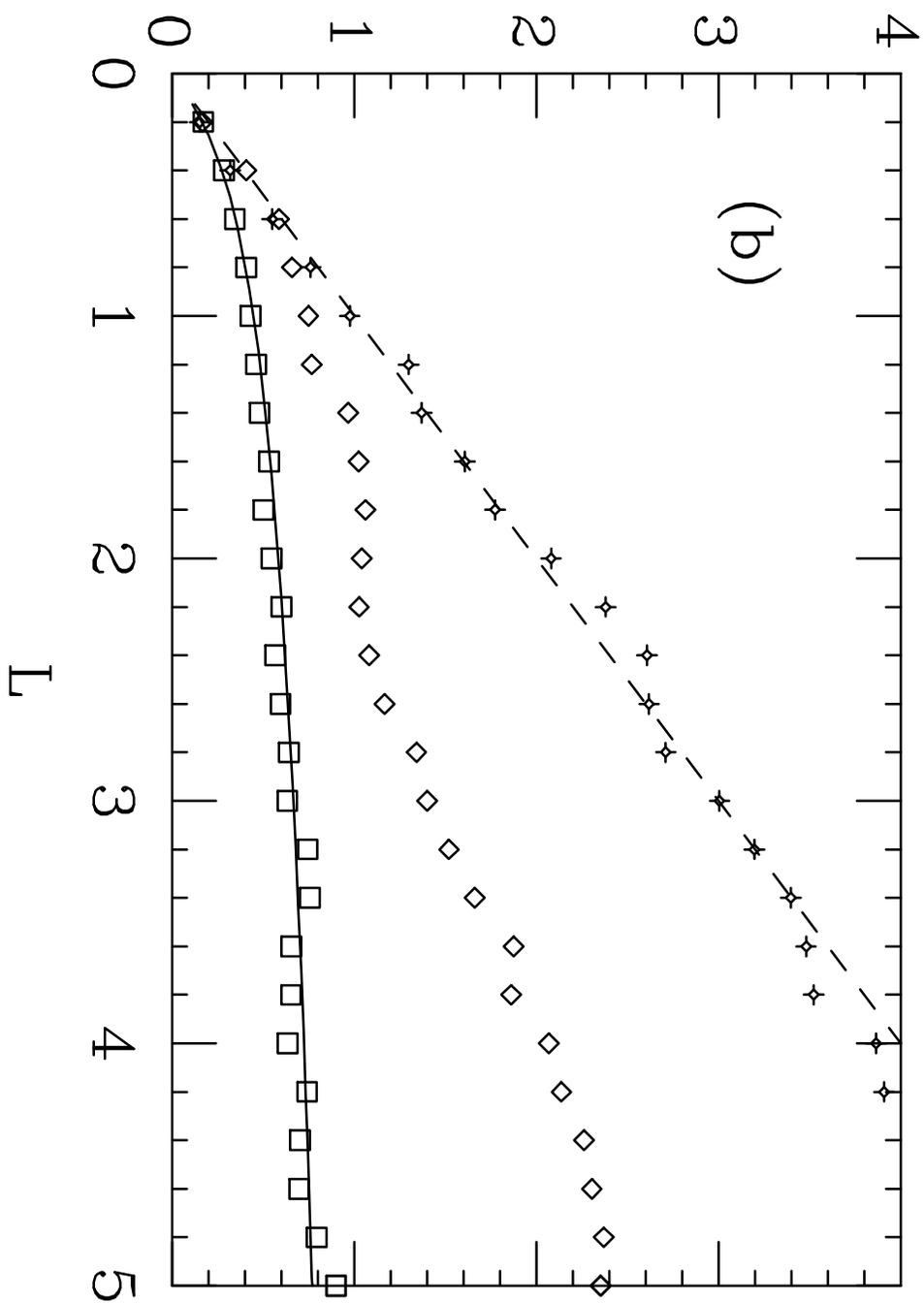
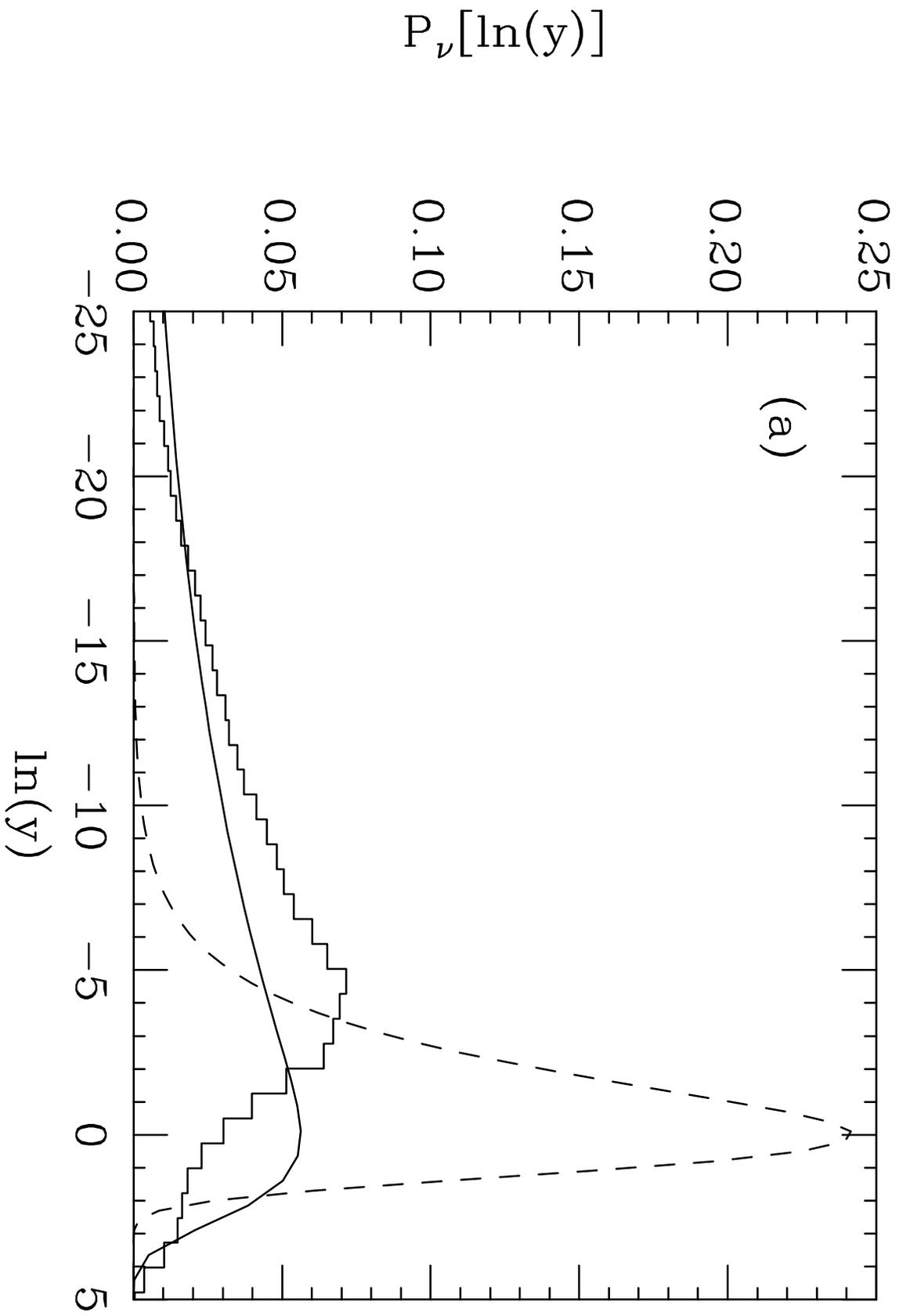
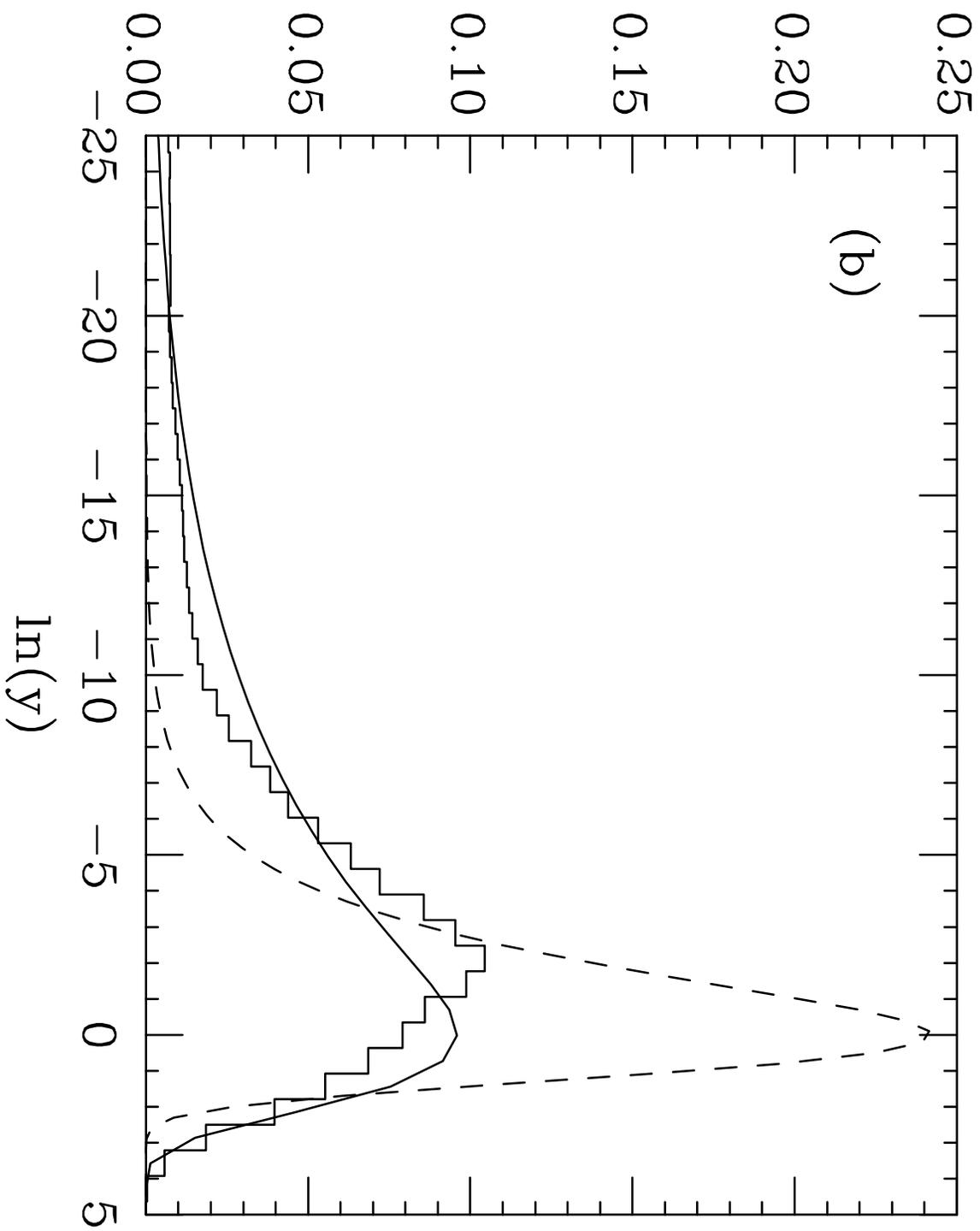
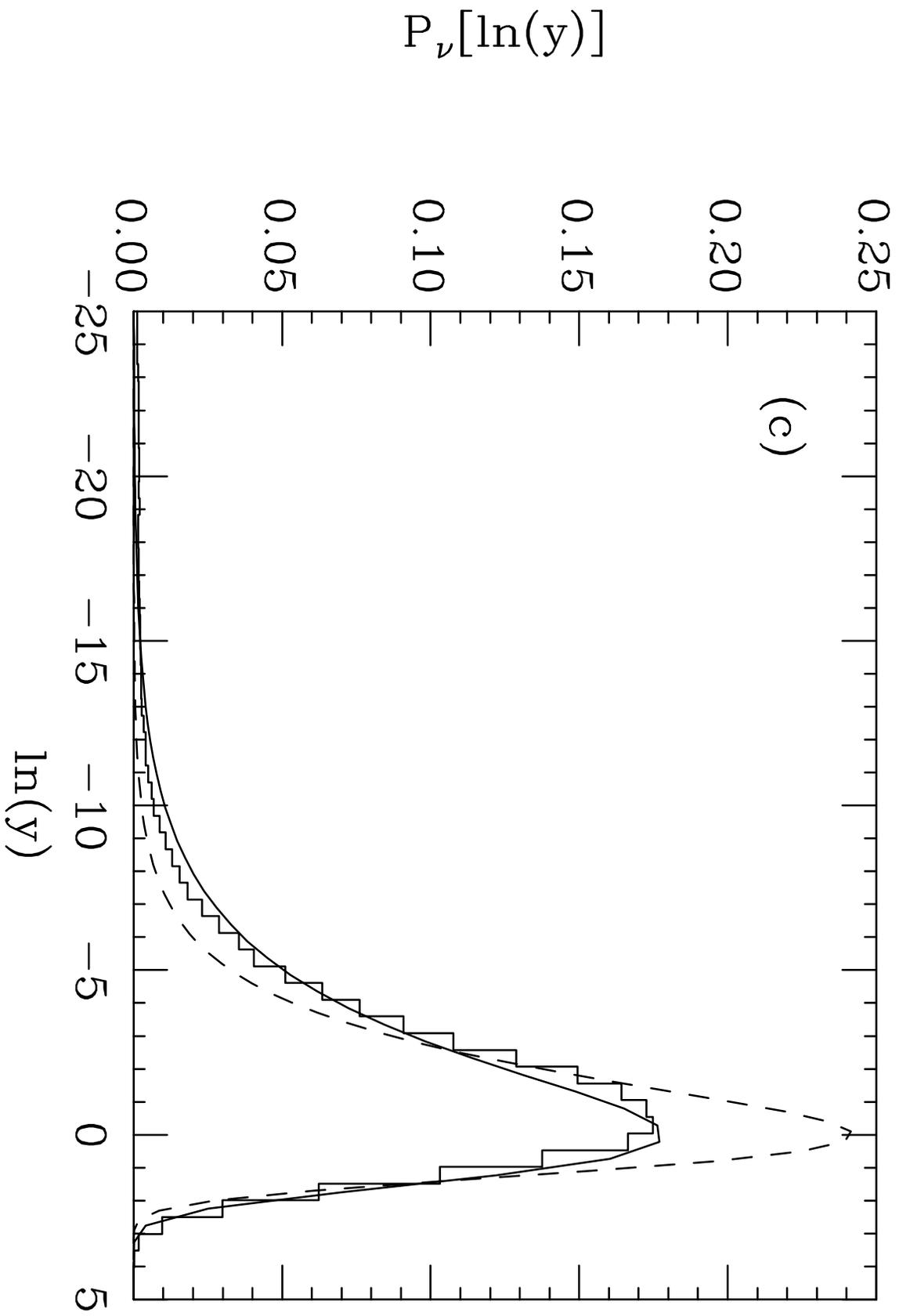
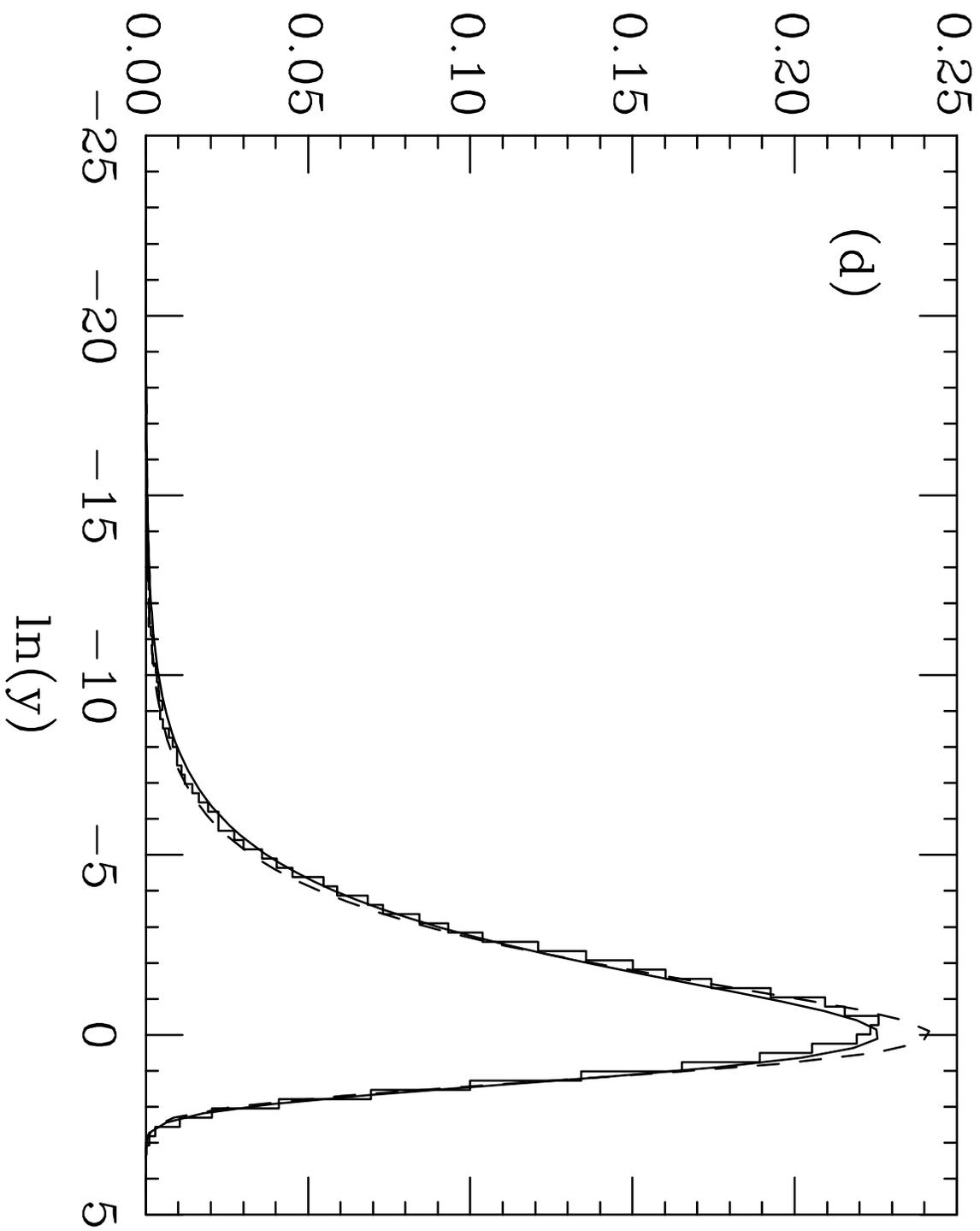
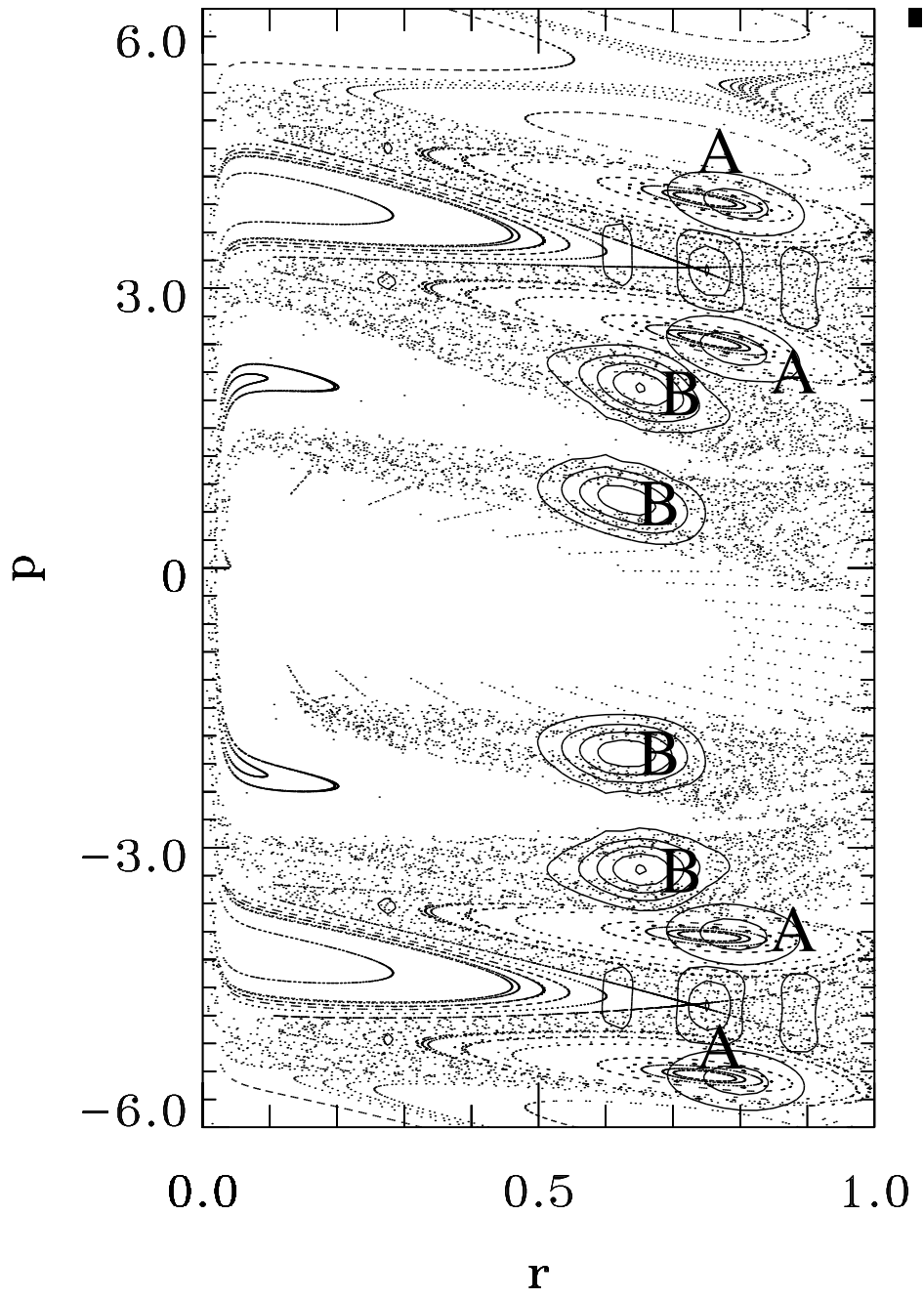
\end{document}